\documentclass[aps,preprint,nofootinbib,epsfig]{revtex4}

\usepackage{epsfig}
\usepackage{mathrsfs}
\usepackage{amsmath}
\usepackage{amssymb}
\usepackage{color}
\usepackage{graphicx}
\usepackage{dcolumn}
\usepackage{bm}

\def\refb#1{(\ref{#1})}
\newcommand{\be}{\begin{equation}}
\newcommand{\ee}{\end{equation}}
\newcommand{\ba}{\begin{eqnarray}}
\newcommand{\ea}{\end{eqnarray}}

\def\refb#1{(\ref{#1})}

\def\b{{\bf{b}}}

\begin{document}

\title{Cosmic ray interaction event generator Sibyll 2.1}

\author{Eun-Joo Ahn$^{1,2}$, Ralph Engel$^3$, Thomas K.~Gaisser$^2$, Paolo Lipari$^4$, and Todor Stanev$^2$} 
\affiliation{$^1$Center for Particle Astrophysics, Fermi National Accelerator Laboratory, Batavia, IL 60510-0500, USA} 
\affiliation{$^2$Bartol Research Institute, Department for Physics and Astronomy, University of Delaware, Newark, DE 19716, USA} 
\affiliation{$^3$ Forschungszentrum Karlsruhe, Institut f\"{u}r Kernphysik, Postfach 3640, 76021 Karlsruhe, Germany} 
\affiliation{$^4$ INFN sezione Roma ``La Sapienza'', Dipartimento di Fisica Universit\'{a} di Roma I, Piazzale Aldo Moro 2, I-00185 Roma, Italy}

\preprint{FERMILAB-PUB-09-304-A}

\date{\today}

\begin{abstract}

The cosmic ray interaction event generator {\sc sibyll} is widely used in extensive air shower simulations. We describe in detail the properties of {\sc sibyll} 2.1 and the differences with the original version 1.7. The major structural improvements are the possibility to have multiple soft interactions, introduction of new parton density functions, and an improved treatment of diffraction. {\sc sibyll} 2.1 gives better agreement with fixed target and collider data, especially for the inelastic cross sections and multiplicities of secondary particles. Shortcomings and suggestions for future improvements are also discussed.

\end{abstract}

\maketitle 

\section{Introduction}

Cosmic ray interactions in the atmosphere can be regarded as high energy fixed target collisions involving heavy particles. Because of their low intensity, cosmic rays with energies above $10^{15}$ eV can only be studied indirectly through the extensive air showers (EAS) they initiate in the atmosphere. The analysis of EAS relies on air shower Monte Carlo simulations which uses hadronic interaction models. At higher energies, where the cosmic ray energy is beyond the reach of man-made accelerators, hadronic interaction properties have to be extrapolated. The difficulties in the extrapolation are augmented by the fact that, while the forward region contains most of the energetics and is important for shower development, most of the accelerator measurements are made in the central region. 

The event generator {\sc sibyll} \cite{Fletcher:1994bd} is intended for air shower cascade simulations. It is a relatively simple model that is able to reproduce many features of hadronic interactions in fixed target and collider experiments. {\sc sibyll} is based on the dual parton model (DPM)~\cite{Capella:1977me,Capella:1981xr,Capella:1992yb}, the Lund Monte Carlo algorithms~\cite{Bengtsson:1987kr, Sjostrand:1987xj}, and the minijet model~\cite{Gaisser:1984pg, Pancheri:1985ix, Durand:1987, Durand:1988cr}. The hard interaction cross section is calculated according to the minijet model. For hadron-nucleus interactions, the interaction probability for each nucleon inside the nucleus is calculated based on the impact parameter distribution. The total interaction cross section is calculated using the Glauber scattering theory~\cite{Glauber:1970jm}. For a nucleus-nucleus interaction the semisuperposition model~\cite{Engel:1992vf} is used to determine the point of first interaction for the nucleons of the projectile nucleus. The fragmentation region is emphasized as appropriate for air shower simulations. Versions 1.6 and 1.7 of {\sc sibyll} have been released and used since the early 1990s. The only difference between the two is that version 1.7 can have neutral pion interactions, which is important only for air showers above $10^{19}$ eV because at lower energy all neutral pions decay before they interact.

Several shortcomings of {\sc sibyll} 1.6 and 1.7 have been identified over the years, such as (1) the total proton-proton cross section calculated with the parton structure functions rose faster than what the experimental measurements indicate; (2) multiplicity fluctuations and average charged particle multiplicity are too small at high energy; (3) diffractive events did not agree well enough with the available data sets. For these reasons the event generator was modified and has been available as {\sc sibyll} 2.1~\cite{Engel:1999db} since 1999. 

The most important changes in version 2.1 are in the description of soft interactions and diffraction dissociation. In order to allow multiple soft interactions, the eikonal for the soft interaction is described using Regge theory, whereas in version 1.7 the eikonal for the soft interactions was energy independent and had the same $b$ dependence ($b$ is the impact parameter) as used for hard interactions. While in version 1.7 the cross section for diffraction dissociation is parametrized independently of the eikonal model, a two-channel eikonal model based on the Good-Walker model~\cite{Good:1960ba, Fletcher:1994hv} is used in {\sc sibyll} 2.1. In addition, low- and high-mass diffraction dissociation are treated separately in the new version. However, it should be kept in mind that diffraction dissociation is still not satisfactorily understood. The parton structure functions have been updated to agree with the HERA results. Other parameters were retuned with updated values as well. The multiple soft interaction and new parton densities give larger multiplicity at high energies and better agreement with data. The multiplicity distribution has been improved by implementing better the effect of diffraction dissociation.

The aim of this paper is to describe the current 2.1 version of {\sc sibyll} to make a reference of the implemented physics models and ideas available. We will outline the overall structure and improvements made, within details of the soft interactions and diffraction dissociation. We compare {\sc sibyll} with fixed target and collider data, and we show how it performs in air shower simulations. Finally, we list some remaining shortcomings of {\sc sibyll} 2.1 and outline how they can be improved.

\section{Hadron-hadron interaction}

\subsection{Basic DPM picture}

{\sc sibyll} 2.1 retains the DPM picture. In the DPM picture, a nucleon consists of a quark ($q$, color triplet) and diquark ($qq$, color antitriplet). Soft gluons are exchanged in an interaction and the color field gets reorganized. The projectile quark (diquark) combines with the target diquark (quark) to form two strings. Each string fragments separately following the Lund string fragmentation model~\cite{Sjostrand:1987xj}.

The fractional energy $x$ of the quark $f_q(x)$ is chosen from a distribution of
\be
\displaystyle
f_q(x) ~=~ \frac{(1 \,-\, x)^\alpha}{(x^2 \,+\, \mu^2/s)^{1/4}} \ ,
\label{eq:valences}
\ee
where $\alpha = 3.0$ and $\mu = 0.35$ GeV is the effective quark mass. The diquark energy fraction is then $f_{qq}(x) = 1 - f_q(x)$. If particles 1, 2 collide to form strings $a$ and $b$, the energy and momentum of the strings are as follows
\ba
E_a ~=~ \frac{\sqrt{s}}{2} \left( x_{1,q} \,+\, x_{2,qq} \right) &,&~~ 
E_b ~=~ \frac{\sqrt{s}}{2} \left( x_{1,qq} \,+\, x_{2,q} \right) \\
p_a ~=~ \frac{\sqrt{s}}{2} \left( x_{1,q} \,-\, x_{2,qq} \right) &,&~~ 
p_b ~=~ \frac{\sqrt{s}}{2} \left( x_{1,qq} \,-\, x_{2,q} \right) 
\ea
To fragment the string, a $q$-$\bar{q}$ pair or $qq$-$\overline{qq}$ pair is generated at one of the randomly chosen ends of the string. The new flavor combines with the existing one to form a hadron, and the remaining (anti)flavor becomes the new end. A primordial $p_T$ of equal magnitude and opposite signs is assigned to the pairs, with a Gaussian distribution where the mean is energy dependent
\be
\langle p_T \rangle ~=~ \left[ p_0 \,+\, 0.08 \log_{10} \left(
    \frac{\sqrt{s}}{30 \, \rm{GeV}} \right)\right] {\rm GeV/c} \ ,
\label{eq:softpt}
\ee
with $p_0 = 0.3$ ($u,d$), 0.45 ($s$), 0.6 ($qq$) and $\sqrt{s}$ being the c.m.\ energy of the hadron-hadron interaction. The energy fraction of each new particle follows the Lund fragmentation function
\be
f(z) ~=~ \frac{(1-z)^a}{z} \, \exp \left[ \frac{-b \, m_T^2}{z} \right] \; ,
\ee
where $a = 0.5$ and $b = 0.8$, $m_T = \sqrt{m^2 + p_T^2}$ is the transverse mass, and $z$ is the fraction of the new particle energy with respect to the parent quark or diquark. The fragmentation process continues until the remaining string mass is less than a ``threshold mass.'' The threshold mass here is defined as the quark masses of the string ends plus the quark/diquark pair mass plus ($1.1 \pm 0.2$) GeV.  The string finishes the fragmentation by forming two final hadrons. 

\subsection{Hard interactions and minijets}

Already in the range of collider energies, at $\sqrt{s} \sim$ 100 GeV, the original DPM picture of just two strings cannot explain what is observed, namely:
\begin{enumerate}
\item high multiplicity;
\vspace*{-2mm}
\item increase of mean $p_T$; 
\vspace*{-2mm}
\item high $p_T$ jets;
\vspace*{-2mm}
\item rise of central rapidity density.
\end{enumerate}
These new features can be interpreted as the emergence of hard interactions which become prominent as energy increases, in the form of minijets. Minijets are described with perturbative QCD, where partons from the colliding hadrons experience hard scattering. Minijets have a transverse momentum larger than some momentum transfer scale $p_T^{\rm min} \gg \Lambda_{QCD}$, where perturbative calculation holds, but smaller than a typically reconstructed collider jet. The minijet formalism  described below is based on Refs.~\cite{Gaisser:1988ra, Fletcher:1994bd} with modifications made in the new version.

Minijets are perceived as part of the hard interaction described by perturbative QCD. The cross section is calculated within the QCD-improved parton model in leading order is
\begin{multline}
  \displaystyle \sigma_{\rm QCD}(s, p_T^{\rm min}) ~=~ K \, \int d x_1 \,
  \int d x_2 \,
  \int d p_T \\
  \times \, \sum_{i,j,k,l} \frac{1}{1 \,+\, \delta_{k,l}} \,
  f_{a,i}(x_1,Q^2) \, f_{b,j}(x_2,Q^2) \, \frac{d \sigma_{\rm
      QCD}^{i,j \to k,l}(\hat{s}, \hat{t})}{d p_T} \,
  \Theta(p_T \,-\, p_T^{\rm min}) \ ,
\label{eq:sigqcd}
\end{multline}
where $f_{a,i}(x_1,Q^2)$ and $f_{b,j}(x_2,Q^2)$ are the parton distribution functions of parton $i$ ($j$) in particle $a$ ($b$). The transverse momentum of the scattered partons is denoted by $p_T$. The calculation is done for four light flavors. Higher order corrections are accounted for by setting the factor $K$=2 and the factorization scale $Q^2 = p_T^2$. {\sc sibyll} 1.7 used parton densities of Ref.~\cite{Eichten:1984eu} (EHLQ), where the gluon density is extrapolated as $g(x) \sim 1/x$ at small $x$. Data from HERA~\cite{Adloff:1997mf, Breitweg:1997hz} suggest a steeper increase at low $x$. {\sc sibyll} 2.1 uses parton densities of Ref.~\cite{Gluck:1994uf, Gluck:1998xa} (GRV) which scales the gluon density as $1/x^{1 + \Delta}$ with $\Delta = 0.3 - 0.4$. As in the previous version, Eq.~\refb{eq:sigqcd} has been calculated separately and is included in the code in tabular form.

The change in the low-$x$ region affects the minijet cross section substantially at high energies. The cross section cannot rise without limit at high energies~\cite{Gribov:1984tu, Levin:1990gg}. If the number of gluons times the transverse resolution scale of hard interaction ($\sim 1/p_T^2$) becomes comparable to the proton size, nonlinear effects, possibly saturation, cannot be neglected. Another factor to take into account is the use of collinear factorization approximation in calculating minijet cross sections, where the transverse momenta of the incoming partons ($i,j$) should always be smaller than the transverse momenta of the scattered partons ($k,l$). This approximation is used to sum the parton densities and only the leading term $\ln(p_T^2)$ is considered. The collinear factorization approximation breaks down for $\ln(1/x) \gg \ln(p_T^2)$. The $\ln(1/x)$ term becomes important at high energies and needs to be taken into account~\cite{Lipatov:1996ts}. In order to restrict the calculation of the minijet cross section to the region of phase space where the QCD-improved parton model is expected to be reliable, the following transverse momentum cut is applied
\be
\displaystyle
p_T^{\rm min}(s) ~=~ p_T^0 \,+\, \Lambda \exp \left[ c \, \sqrt{\ln (s / \textrm{GeV}^2)} \right] \ ,
\label{eq:ptmin}
\ee 
where $p_T^0 = 1$ GeV, $\Lambda = 0.065$ GeV, $c=0.9$. This parametrization follows from the geometric saturation condition~\cite{Levin:1990gg}
\begin{equation}
\frac{\alpha_s(p_T^2)}{p_T^2} \cdot  x g(x,p_T^2) ~\le~ \pi R_p^2 \ ,
\label{eq:saturation}
\end{equation}
where $\alpha_s$ is the strong coupling constant, $g(x,p_T^2)$ the gluon density, and $R_p$ is the effective radius of a proton in transverse space. The scale $Q^2 = p_T^2$ is assumed. In the limit $\ln(1/x)$,~$\ln(Q^2/\Lambda^2) \rightarrow \infty$ (double leading-logarithmic approximation) the steeply rising gluon density $g(x,Q^2)$ can be written 
\be
x g(x,Q^2) \sim \exp\left[ \frac{48}{11-\frac{2}{3}n_f}
\ln \frac{\ln \frac{Q^2}{\Lambda^2}}{\ln \frac{Q_0^2}{\Lambda^2}} \ln
\frac{1}{x} \right]^{\frac{1}{2}}
\sim \frac{1}{x^{0.4}}, 
\label{eq:double-log}
\ee 
with $\Lambda$ being the QCD renormalization scale and $n_f$ is the number of quark flavors. The functional form of Eq.~\refb{eq:ptmin} follows from inserting Eq.~\refb{eq:double-log} in Eq.~\refb{eq:saturation}, however, the parameters in Eq.~\refb{eq:ptmin} cannot be derived directly from first principles.

\begin{figure}
\includegraphics[width=0.5\textwidth]{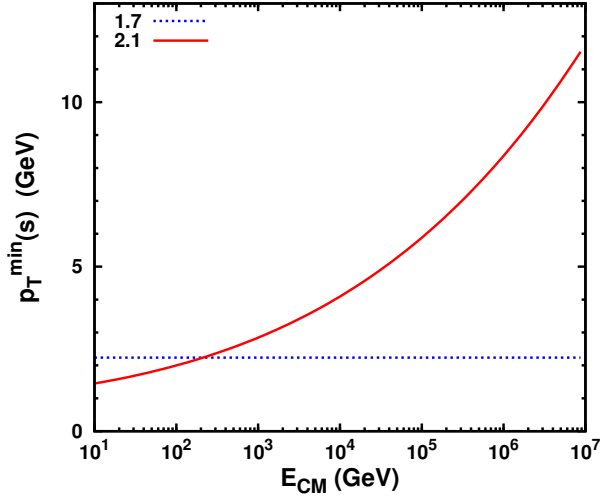}
\caption{Minimum transverse momentum ($p_T^{\rm min}$) required for the collision to qualify as a hard scattering. {\sc sibyll} 1.7 had a constant minimum at $\sqrt{5}$ GeV, whereas this has been modified to change with $\sqrt{s}$ in version 2.1.}
\label{fig:ptmin}
\end{figure}

{\sc sibyll} 1.7 had a constant cutoff at $p_T^{\rm min} = \sqrt{5}$\,GeV. Using modern parton density parametrizations one cannot obtain a satisfactory description of the proton-proton/antiproton cross sections with a transverse momentum cutoff that is energy-independent. The $p_T^{\rm min}$ values for both versions are shown in Fig.~\ref{fig:ptmin}.

The minijet cross section quickly rises to exceed the total cross section. This is interpreted as the collision forming more than one minijet. The average number of hard interactions $n_{\rm hard}$ occurring at energy $s$ and at impact parameter $b$ is~\cite{Durand:1987}
\be
n_{\rm hard}(b,s) ~=~ A(b) \, \sigma_{\rm QCD}(s) \ ,
\label{eq:nmean}
\ee
where $A(b)$ is the profile function for the hadron-hadron collision. The baryon and meson profile functions follow those of Refs.~\cite{Durand:1988cr, Durand:1990qa} and are given in Appendix \ref{app:hard}. Following the convention given in Ref.~\cite{Block:1984ru}, where the basic equations are in Appendix \ref{app:conv}, the inelastic cross section is
\be
\displaystyle
\sigma_{\rm inel} ~=~ \int d^2 b \left[ 1 \,-\, e^{-2 \chi(b,s)} \right] \ ,
\label{eq:siginel}
\ee
where the eikonal is
\be
\chi(b,s) ~=~ \chi_{\rm hard}(b,s) \,+\, \chi_{\rm soft}(b,s) ~=~
\frac{1}{2} n_{\rm hard}(b,s) \,+\, \frac{1}{2} n_{\rm soft}(b,s) \ .
\label{eq:eikonal}
\ee
The number of soft interactions is defined analogously to the hard one $n_{\rm soft}(b,s) = A_{\rm soft}(b) \sigma_{\rm soft}(s)$. The soft part of the eikonal is discussed in the following subsection.

The hard part of the eikonal is interpreted as having a probability of $\exp[-n_{\rm hard}(b,s)]$ for no minijet production at energy $s$ and impact parameter $b$.  Equation \refb{eq:siginel} can be reorganized as
\ba
\displaystyle
\sigma_{\rm inel} &=& \int d^2b \, \left\{ 1 \,-\, e^{-n_{\rm hard}(b,s)} \,+\, 
e^{-n_{\rm hard}(b,s)} \,-\, e^{-n_{\rm hard}(b,s)} \, e^{-n_{\rm soft}(b,s)} \right\} \nonumber \\ 
&=& \sum_{N=1}^{\infty} \sigma_N ~+~ \int d^2b \, 
e^{-n_{\rm hard}(b,s)} \, \left[ 1 - e^{-n_{\rm soft}(b,s)} \right] \ ,
\ea
where
\be
\sigma_N ~=~ \int d^2b \, \frac{n_{\rm hard}(b,s)^N}{N !} \,
e^{-n_{\rm hard}(b,s)}
\ee
is the cross section for production of $N$ pairs of minijet. This interpretation follows from the Abramovski-Gribov-Kancheli \cite{Abramovsky:1973fm} cutting rules, where $\sigma_N$ is the term with exactly $N$ cut parton ladders, summed over all uncut ladders~\cite{TerMartirosyan73}. The probability distribution for obtaining $N$ minijet pairs is
\be
P_N ~=~ \frac{\sigma_{N}}{\sigma_{\rm inel}} \ ,
\label{eq:pn}
\ee
 and the mean number of minijet pairs produced per interaction is
\be
\langle N \rangle ~=~ \sum_{N=0}^{\infty} N \, P_N ~=~ \frac{\sigma_{\rm QCD}}{\sigma_{\rm inel}} \ .
\label{eq:nminimean}
\ee

\begin{figure}
\includegraphics[width=0.49\textwidth]{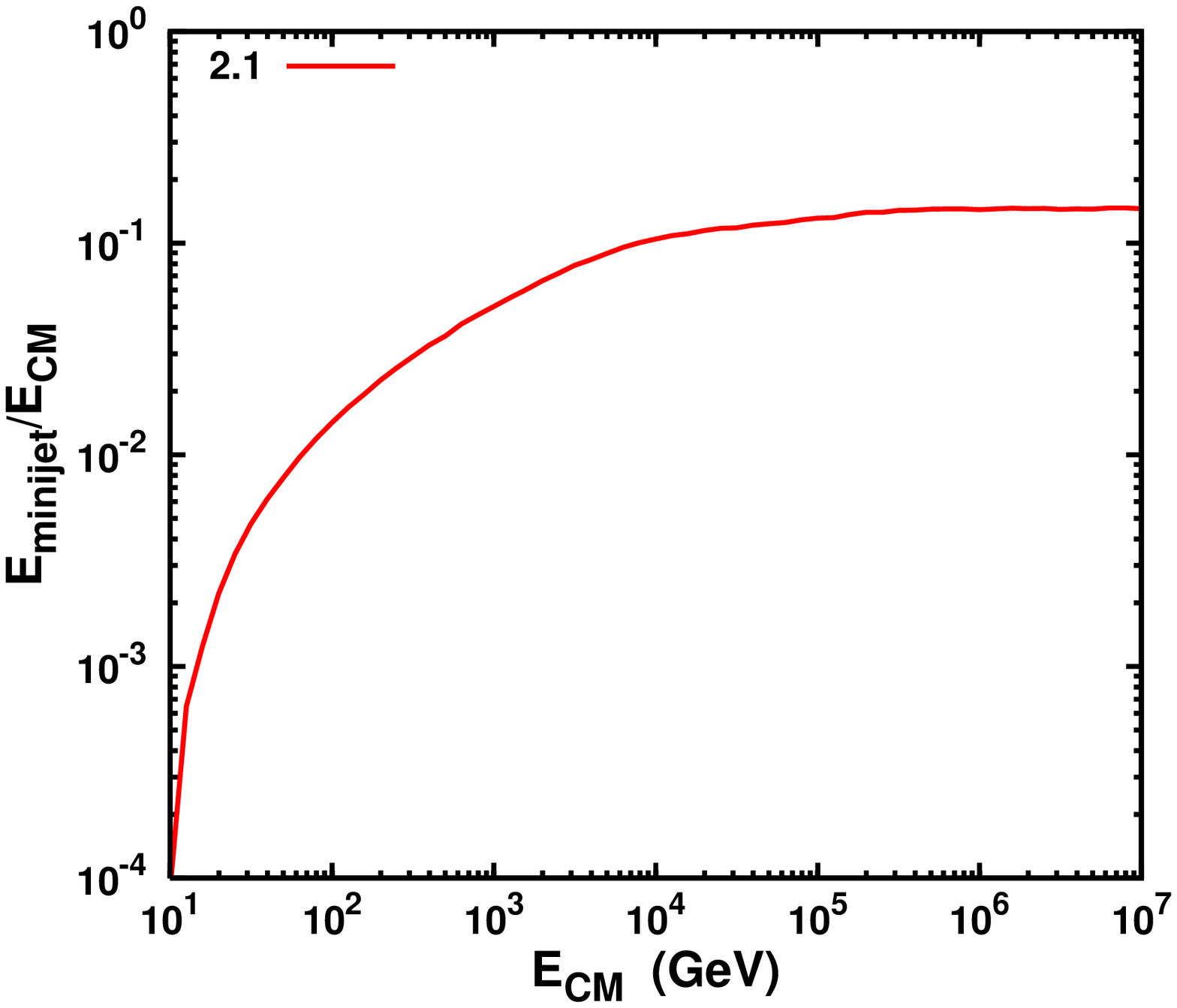}
\includegraphics[width=0.49\textwidth]{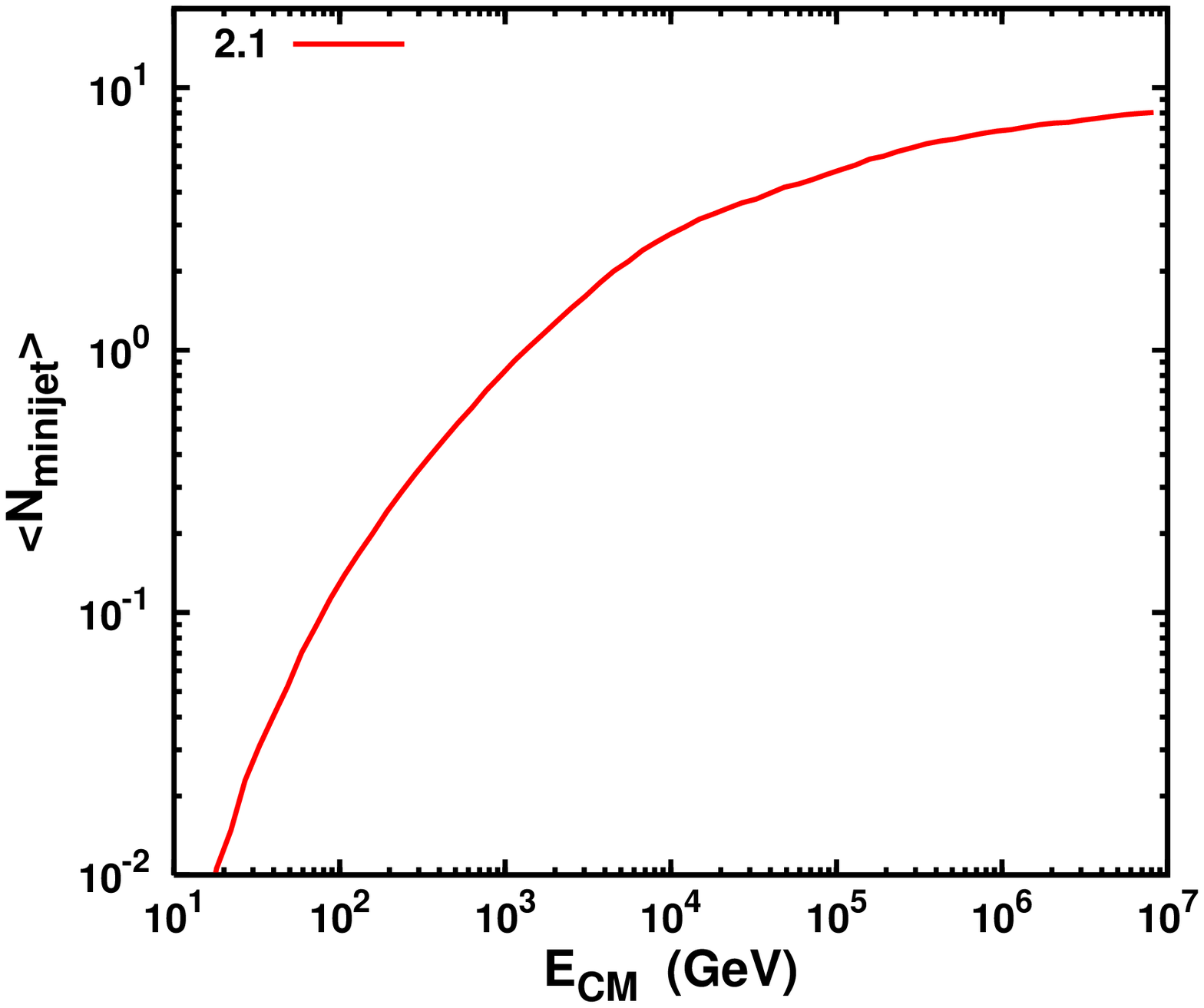}
\caption{Minijet production, with the energy fraction carried by the minijets (left panel) and the average number of minijets produced (right panel) over a range of $\sqrt{s}$.}
\label{fig:minijets}
\end{figure}

The contribution of minijets to the overall particle production for a $p$-$p$ collision is shown in Fig.~\ref{fig:minijets}, where the energy fraction carried by the minijets and mean number of minijets produced are shown as a function of c.m.\ energy.
Minijets start becoming important at $\sqrt{s} \approx 1000$ GeV.

Each minijet pair is treated as two strings stretched between two gluons. In order to fragment the jets, a $q$-$\bar{q}$ pair is generated at each end, and a leading particle at each end is created. Then the string fragments in the standard way. The fraction of energy going into the minijet from each hadron 1, 2 ($x_1$ an $x_2$) is obtained by selecting $x$ from the effective parton density function \cite{Combridge:1983jn}
\be 
\displaystyle 
f(x) ~=~ g(x) \,+\, \frac{4}{9} \left[ q(x) \,+\, \bar{q} (x) \right] \ .
 \ee 
The code uses the approximation that for small transfer momentum the cross sections for $g$-$g$, $q$-$g$ and $q$-$q$ scattering are proportional to $t^{-2}$ and are in ratio $1: 4/9: (4/9)^2$. The transverse momentum follows 
\be 
\frac{d \sigma}{d \hat{t}} ~\propto~ \frac{1}{\hat{t}^2} \ , 
\ee 
where $\hat{t}$ is the four-momentum transferred squared Mandelstam variable, and $\hat{t} > \left( p_T^{\rm min} \right)^2$. We emphasize that the full parton structure functions of the $u,~d,~s,~c$ quarks and gluon are used for the calculation of the hard cross section. The above approximation is made only when sampling partonic final states and the parton density is parametrized in this simple way by adding quarks and gluons with the approximate weights.

\subsection{Soft interactions}

{\sc sibyll} 1.7 has a very simple, energy-independent form of soft contribution to the eikonal $\chi_{\rm soft} = \frac{1}{2}  C A(b)$, having an impact parameter profile function identical to that of the hard counterpart, and $C = 123$ GeV$^{-2}$ is chosen to reproduce the low energy inelastic cross section of 32 mb. Only one soft interaction is permitted, and the hard-soft interaction division was energy-independent at $p_T = \sqrt{5}$ GeV. The energy left over after the production of minijets was shared by two strings connecting the valence quarks of the projectile and target.

In version 2.1, the energy-dependent $p_T^{\rm min}(s)$ allows a larger range of phase space for soft interactions. The eikonal form of $\chi_{\rm soft} = \frac{1}{2} A_{\rm soft}(b) \sigma_{\rm soft}(s)$ is kept. We adopt some aspects of Regge theory in order to accommodate multiple soft interactions. Inspired by Regge theory \cite{Collins77} the energy dependence of $\sigma_{\rm soft}$ is taken as sum of two power laws,
 one term for Pomeron exchange and another term for Reggeon exchange \cite{Donnachie:1992ny}
\be
\sigma_{\rm soft}(s) ~=~ X \, \left( \frac{s}{s_0}
\right)^{\Delta_{\rm eff}} \,+\, Y \, \left( \frac{s}{s_0} \right)^{-\epsilon} \ ,
\label{eq:sig-dl}
\ee
The index $\epsilon$ for Reggeon exchange at low energy is expected to be very similar to the one found in fits by Donnachie and Landshoff~\cite{Donnachie:1992ny}. The parameter $\Delta_{\rm eff}$, in contrast, depends on the subdivision of the Pomeron term into soft and hard contributions and is hence a function of the transverse momentum cutoff \refb{eq:ptmin}. Here we implicitly assume that minijets form the hard part of the Pomeron \cite{Capella:1986cm}.

The parameters $X$, $Y$ and $\epsilon$ and $\Delta_{\rm eff}$ are determined by fitting the measured total, elastic and inelastic cross sections for $p$-$p$ and $p$-$\bar p$ interactions. Based on the GRV parton densities \cite{Gluck:1994uf, Gluck:1998xa} $\epsilon \approx 0.4$ and $\Delta_{\rm eff} \approx 0.0245$ are found.

In the following we describe soft interaction by extrapolating the picture of hard interactions into the domain of low momentum transfer, in which this picture cannot be justified within perturbative QCD. Only comparison of the corresponding model predictions with data can later prove this description as useful.

\begin{figure}
\unitlength1mm
\begin{picture}(60,52)
\put(0,0){\epsfig{figure=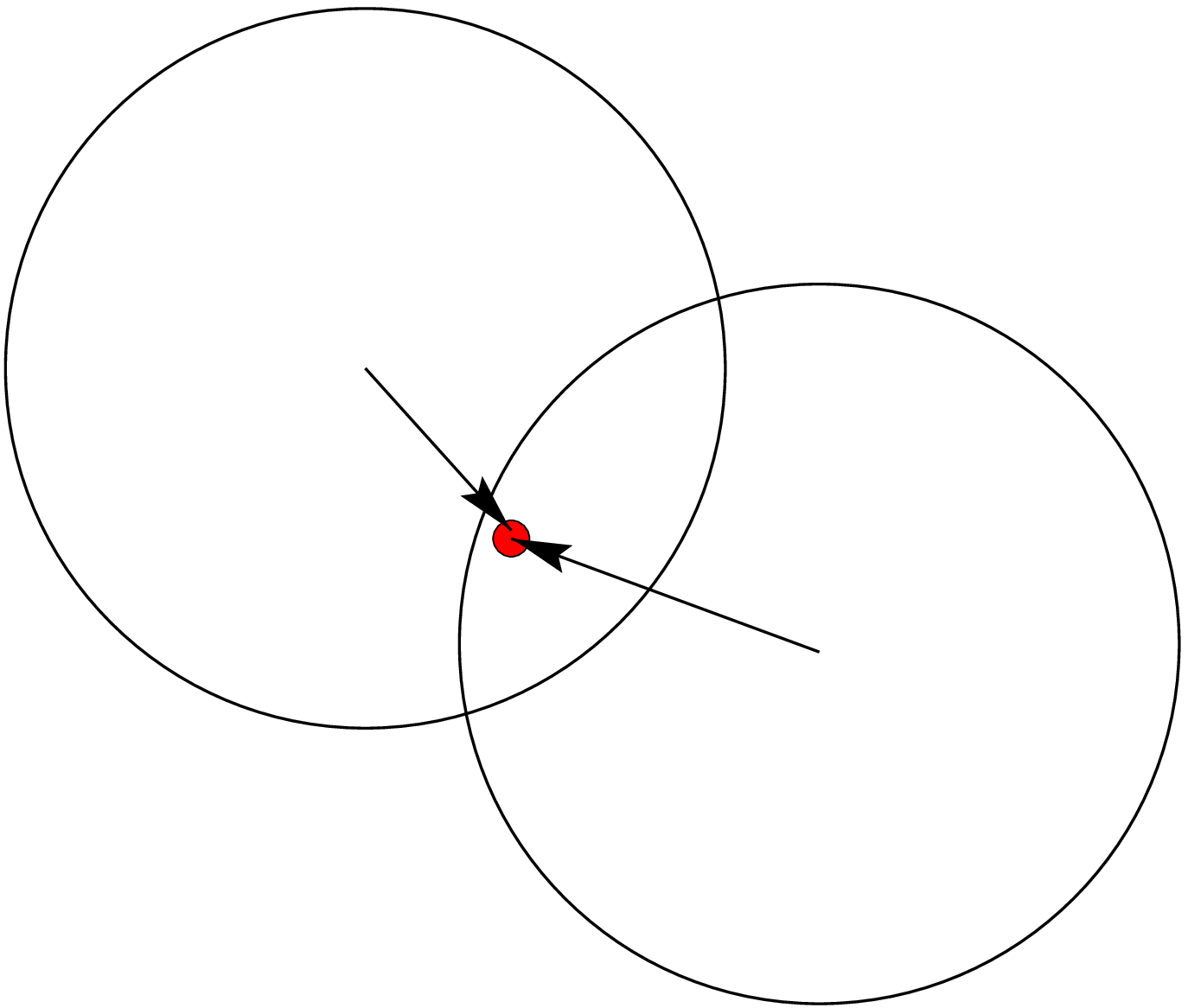,width=6cm}}
\put(15,27){$\b_1$}
\put(35,15){$\b_2$}
\end{picture}
\hspace*{2.5cm}
\begin{picture}(60,52)
\put(0,0){\epsfig{figure=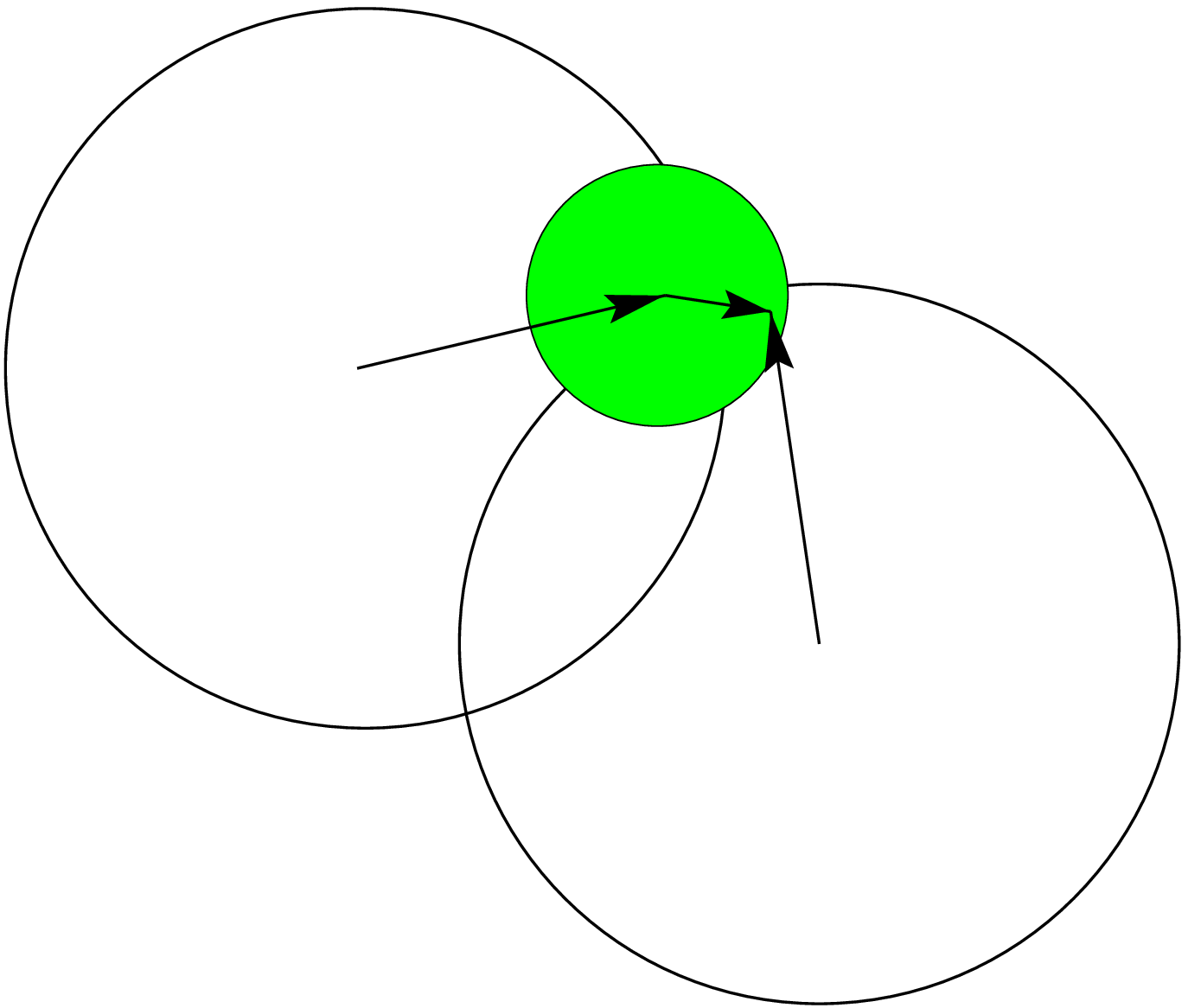,width=6cm}}
\put(20,36){$\b_1$}
\put(44,26){$\b_2$}
\put(39,43){$\b_3$}
\end{picture}
\caption{Geometric interpretation of a hard interaction (left) and soft interaction (right).}
\label{fig:hs}
\end{figure}

While hard interactions are approximately pointlike in character, soft interactions involve a larger transverse interaction area. The low $p_T$ of the partons in a soft interaction spreads out the interaction area (from the uncertainty principle $\Delta p \Delta b \sim 1$), as opposed to a high $p_T$ event which localizes the collision to a small region. The geometry of hard and soft interactions in impact parameter space is schematically shown in Fig.~\ref{fig:hs}. 
The energy-dependent increase of the interaction region of soft interactions has to be taken into account in the soft profile function of the hadron-hadron collision
\be
A_{yz}^{\rm soft}(s,\b) \,=\, \int d^2 {\bf{b_1}} \, d^2 {\bf{b_2}} \, d^2
{\bf{b_3}} \, A_y({\bf{b_1}}) \, A_z({\bf{b_2}}) \,
A^{\rm soft}(s,{\bf{b_3}}) \, \delta^{(2)}({\bf{b_1}}-{\bf{b_2}}+{\bf{b_3}}-\b) \ ,
\label{eq:appsoft}
\ee
where the profile function $A_{y/z}$ is analogous to the hard interaction case. The fuzzy area of the right hand diagram in Fig.~\ref{fig:hs} is represented by $A^{\textrm{soft}}(s,{\bf{b_3}})$, which is parametrized as a Gaussian with a energy-dependent width
\be
A^{\rm soft}(s,{\bf{b_3}}) ~=~ \frac{1}{4 \, \pi \, B_s(s)} \, \exp \left[ - \frac{ |\bf{b_3}|^2}{4 \, B_s(s)} \right] \ ,
\label{eq:softgauss}
\ee
with
\be
B_s(s) ~=~ B_0 \,+\, \alpha^\prime(0) \ln \left( \frac{s}{s_0} \right) \ .
\ee
In the limit of $B_s(s) \to 0$, $A^{\textrm{soft}}(s,{\bf{b_3}})$ becomes a delta function and Eq.~\refb{eq:appsoft} becomes equal to the hard profile function.

In order to calculate Eq.~\refb{eq:appsoft}, an exponential form factor which corresponds to a Gaussian shape in transverse space is used for the proton or meson profile function $A_y(b)$ as a first estimate, using data to fit the parameters. For a proton-proton collision, this yields
\be
A_{pp}^{\rm soft}(s,\b) \,=\, \frac{1}{4 \, \pi \, (2 B_p \,+\, B_s(s))} \, \exp \left[- \frac{b^2}{4 \, (2B_p \,+\, B_s(s))} \right] \ ,
\ee
where $B_p$ characterizes the transverse size of a proton and is fitted to data.

For generating the string configurations in inelastic events, the number of soft ($N_s$) and hard ($N_h$) interactions is sampled from (see also \cite{Aurenche92a})
\be
\sigma_{N_s,N_h} ~=~ \int d^2b \, 
\frac{[n_{\rm soft}(b,s)]^{N_s}}{N_s !} 
\frac{[n_{\rm hard}(b,s)]^{N_h}}{N_h !} \, e^{-n_{\rm hard}(b,s)-n_{\rm soft}(b,s)},
\label{eq:prob}
\ee
with the inelastic cross section given by
\be
\sigma_{\rm inel} = \sum_{N_s+N_h\ge 1} \sigma_{N_s,N_h}\ .
\ee
The probability distribution Eq.~\refb{eq:prob} is tabulated during initialization of {\sc sibyll} and later used to draw event configurations.

In analogy to hard interactions, soft interactions are simulated by a pair of gluons which are fragmented the same way as a minijet pair. Thus, there is one valence quark string pair and $n_s-1$ gluon pairs. The only difference is the lower transverse momentum, see Eq.~\refb{eq:softpt}, and the distribution of the momentum fractions of these gluons, which are sampled from an $1/x$ distribution. This distribution corresponds to the one expected for a scenario of saturated gluon density in central collisions. A minimum mass of $m_{\rm soft} = 1$\,GeV is required for strings between sea quarks to regularize the singular part of the distribution and to ensure applicability of string fragmentation. One of the multiple soft interactions always involves the valence quarks and the momentum fractions are then sampled from Eq.~\refb{eq:valences}.

Implementing multiple soft interactions affects the model predictions at intermediate energy, which can be seen, for example, in the inelastic and total cross sections between $\sqrt{s} = 50 - 900$ GeV in Fig.~\ref{fig:sibcross}.

\subsection{Diffraction dissociation}

Diffraction is a collision where there are no quantum numbers exchanged between the colliding particles. A characteristic feature is a large rapidity gap in the final state. Unfortunately diffraction physics is not satisfactorily understood even on the level of phenomenology. A comprehensive description of diffraction can be found in Ref.~\cite{Barone2002}. We give only a brief description of the diffraction model used in the new version and put the relevant equations in Appendix \ref{app:diff}.

In version 1.7, diffraction was considered part of the inelastic, no-minijet event but was not otherwise included within the physics framework. The cross sections $\sigma_{\rm diff}$ were simply 9\% each for forward and backward diffraction and 4\% for double diffraction of the $\sigma_{\rm inel}$ (including minijet production) at 30 GeV, and was assumed to increase with energy as $\sigma_{\rm diff} \propto \ln(s)$, with the diffractive event probability $P_{\rm diff} = \sigma_{\rm diff}/\sigma_{\rm inel}$. With this treatment, however, $\sigma_{\rm diff}$ becomes larger than the cross section for inelastic no-minijet events at high energy. Also, the minijet cross section is a fit to the total inelastic cross section, which includes diffractive events as well. This resulted in an underestimation of minijet production.

{\sc sibyll} 2.1 uses the two-channel eikonal model to incorporate diffraction into the eikonals for the low-mass diffraction dissociation. The procedure is similar to the Good-Walker model~\cite{Good:1960ba} with only a few assumptions made; see also Ref.~\cite{Kaidalov:1979jz}. Only two states are distinguished, a nonexcited state and a generic diffractively excited state (denoted with a $^\star$) which is a superposition of different low-mass diffractive final states. The cross sections and relevant equations are given in Appendix \ref{app:diff}. The high-mass diffraction dissociation is calculated by extracting the corresponding eikonal function from the soft and hard eikonal parts. Including the excited state of projectile and target in the eikonal formalism gave considerable improvement to the multiplicity distribution (see. Sec.~\ref{sec:nch}).

Diffraction dissociation is treated with a strict kinematic cutoff $M_x^2/s < 0.1$, which follows from considerations on coherence and diffractive particle production \cite{Goulianos:1982vk}. The net effect is that the quasielastically scattered protons do not lose more than $\sim$ 20 \% at maximum in diffraction dissociation at this energy.
The diffractively dissociating particle undergoes a phase-space decay if the mass of the excited system is very low. For higher masses, diffracted particles are divided into two valence components of quark-diquark or quark-antiquark that are connected by a color string which subsequently fragments. The string carries the diffracted particle's momentum and quantum numbers and does not create extra $p_T$. The one-string decay threshold is set to $\Delta M = 0.7$ GeV, where $\Delta M$ is the mass difference of the incoming particle and the excited state of it.

Diffractively excited states of a mass of more than 10 GeV are considered as being produced by a Pomeron-hadron interaction. The decay of these states is described with multiple soft and hard interactions by generating a $\pi$-$p$ interaction at $\sqrt{s} = \Delta M$, as motivated by data from the UA4 Collaboration \cite{Bernard:1985kh}.

\subsection{Nucleus interactions}

The physics framework for the hadron-nucleus and nucleus-nucleus interactions remains the same in the new version. The difference in the cross sections between the versions comes from the improvements made to the hadron-hadron interactions. Detailed descriptions can be found in Ref.~\cite{Fletcher:1994bd} for the hadron-nucleus interaction and Ref.~\cite{Engel:1992vf} for the nucleus-nucleus interaction. Here, we summarize the basic concepts for the sake of completeness.

\begin{figure}
\includegraphics[width=0.49\textwidth]{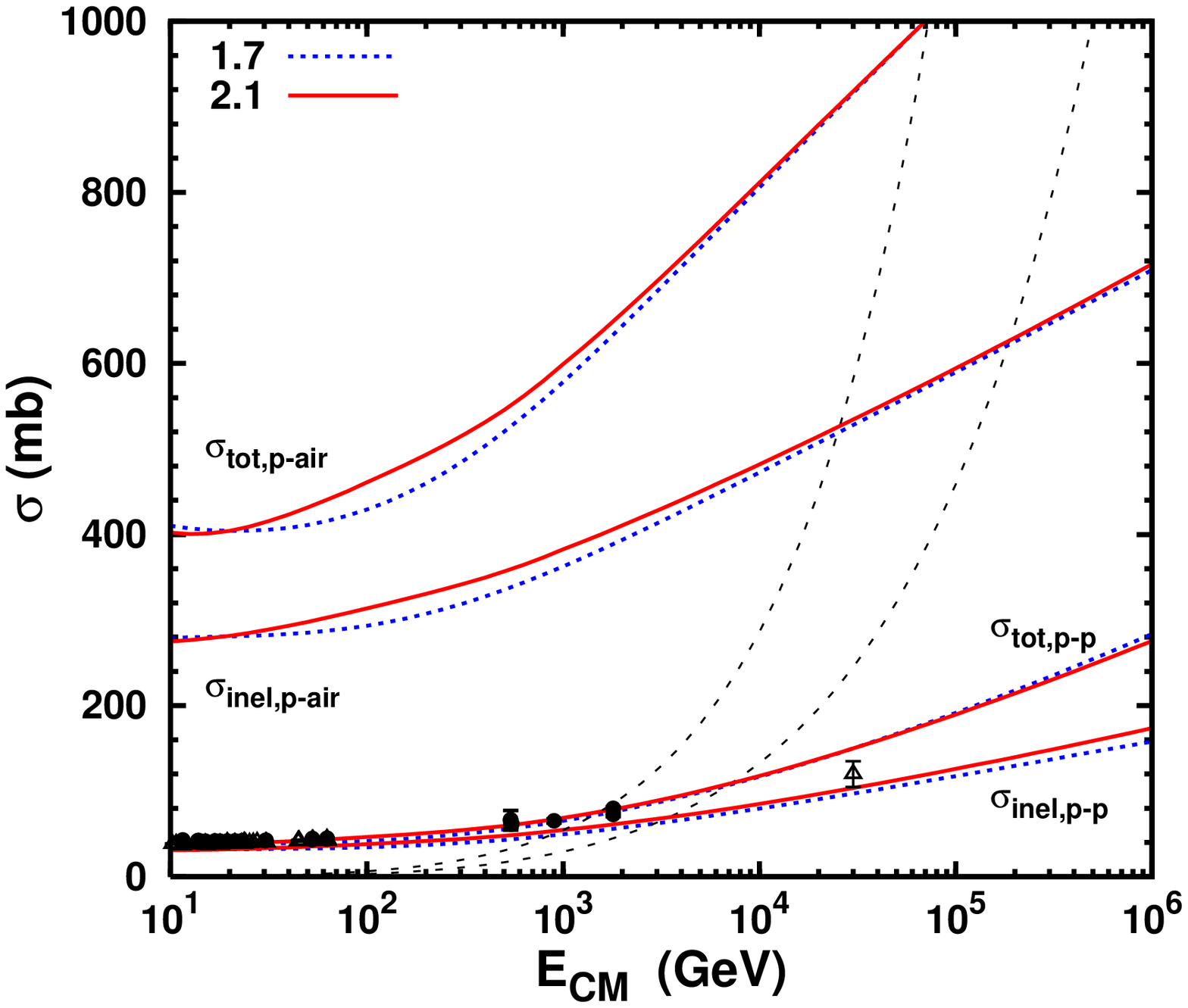}
\includegraphics[width=0.49\textwidth]{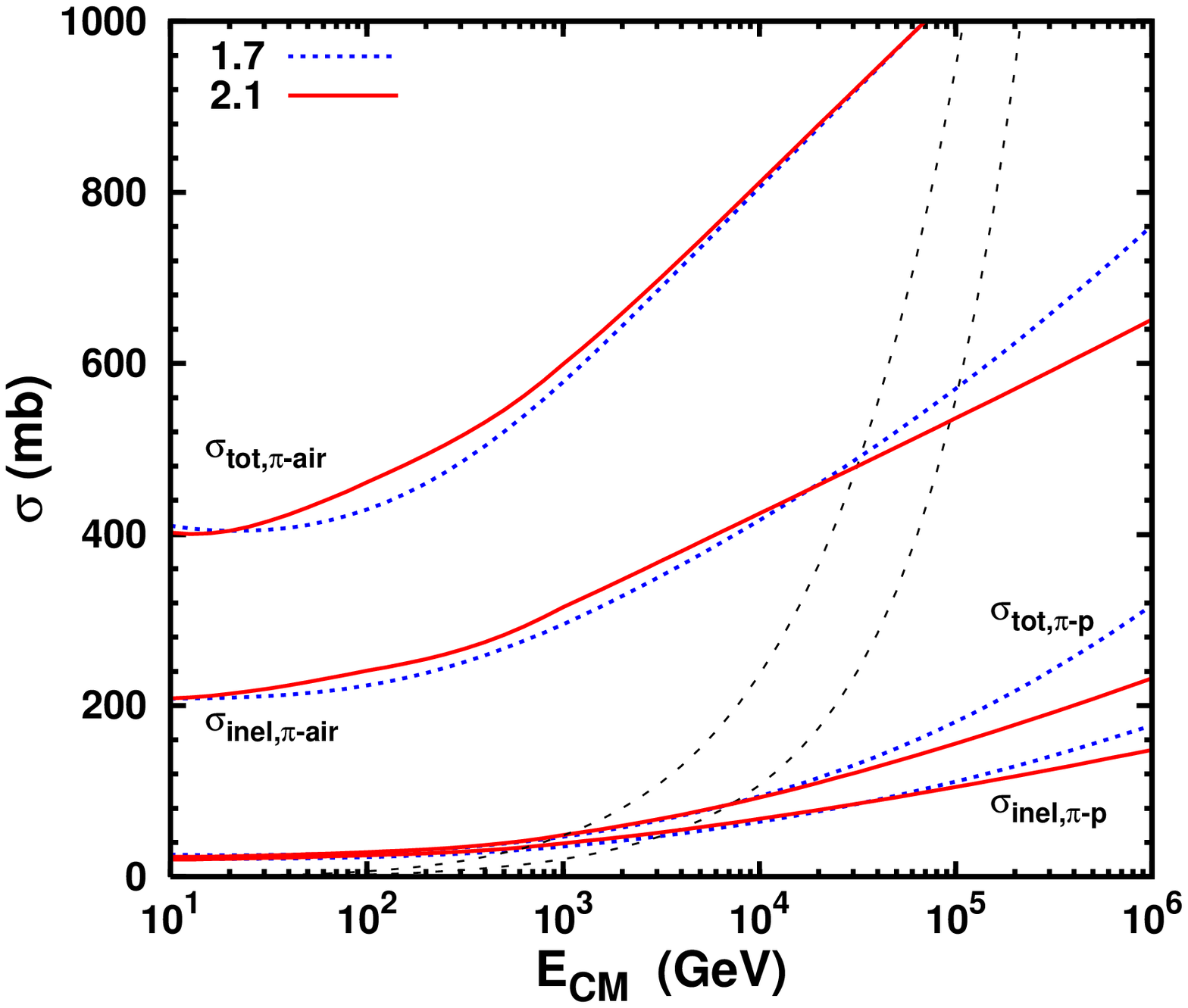}
\caption{The total and inelastic cross section of $p$-air, $p$-$p$ (left panel) and $\pi$-air, $\pi$-$p$ (right panel) collision for version 2.1 (red solid lines) and 1.7 (blue dotted lines). The minijet cross section is also shown in dashed lines, where the higher one is version 2.1.}
\label{fig:sibcross}
\end{figure}

The interaction length is calculated from the production cross section. Elastic and quasielastic interactions, where no new particles are produced, do not contribute to the air shower development and are not considered. The production cross section, i.e. contribution to particle creation, for a hadron-nucleus ($hA$) interaction where $A$ is the mass number of the nucleus is 
\be
\sigma_{\rm prod}^{hA} ~=~ \sigma_{\rm tot}^{hA} \,-\, \sigma_{\rm el}^{hA} \,-\, \sigma_{\rm qe}^{hA} \ ,
\ee
with $\sigma_{\rm tot}^{hA}$, $\sigma_{\rm el}^{hA}$, and $\sigma_{\rm qe}^{hA}$ being the total, elastic, and quasielastic cross sections, respectively.
They are calculated within the Glauber model \cite{Glauber:1970jm} from the $p$-$p$, $\pi$-$p$, and $K$-$p$ cross sections.
The inelastic and total cross sections of $p$-$p$, $p$-air and $\pi$-air, $\pi$-$p$ collision are shown in Fig.~\ref{fig:sibcross}. The minijet cross sections are also shown.

In a hadron-nucleus interaction, the number of target nucleons directly participating in the interaction, also known as wounded nucleons, is determined from the production cross section. The mean number $\langle N_w \rangle$ of wounded nucleon per interaction is given by standard Glauber theory. In analogy with Eqs.~\refb{eq:pn} and \refb{eq:nminimean}, one finds
\be
\langle N_w \rangle ~=~ \frac{1}{\sigma_{\rm prod}^{hA}} \sum_{N_w} N_w \sigma_{N_w} ~=~ \frac{A \, \sigma_{\rm inel}^{hp}}{\sigma_{\rm prod}^{hA}} \ ,
\ee
where $\sigma_{N_w}$ is the cross section for interaction with $N_w$ nucleons and $\sigma_{hp}$ is the hadron-nucleon cross section. 

The string model is applied to the fragmentation of the partonic system as well. The target nucleus is seen as $N_w$ pairs of valence $q$-$qq$, and the projectile hadron is viewed as one valence $q$-$\bar{q}$ or $q$-$qq$ pair and $N_w\!-\!1$ sea $q$-$\bar{q}$ pairs. The $N_w$ color-connected partons undergo string fragmentation. Most of the energy is carried by the valence pair string, and the sea strings contribute to giving the proper multiplicity in the target region. This happens because there are more valence quarks on the nucleus side for the sea quarks to couple to. Intranuclear interactions and Fermi momentum inside the target nucleus are ignored in {\sc sibyll}. Intranuclear interactions are greatly suppressed at high energy due to the time needed for a hadron to form as an independent object (formation time). At high energies, minijets are added to the collision, by generating them for each of the $N_w$ wounded nucleons in the same way as in a hadron-hadron interaction.

The nucleus-nucleus interaction is treated with the semisuperposition model~\cite{Engel:1992vf}, which is between the simple superposition model and full Glauber theory \cite{Glauber:1970jm}. The superposition model treats each nucleon of the projectile independently and as a consequence the interaction lengths of the nucleons have an exponential distribution based on the hadron-nucleus cross section. In reality, the nucleus interaction length is very small and a nucleus will interact quickly in the atmosphere. In the semisuperposition model, the number of interacting nucleons in the projectile for each nucleus interaction is determined from Glauber theory, where the remaining spectator nucleons fragment into lighter nuclei. Though the interaction and fragmentation is treated as a nucleon-nucleus interaction, the distribution of these nucleon-subshowers reflects correctly the nucleus-air cross section.

\section{Comparison with experimental data}

Both fixed target and collider experiments give valuable guidance in modeling hadronic interactions. Fixed target experiments provide data for the forward region which are most relevant to cosmic ray interactions, but the energies are relatively low, $E_{\rm lab} \sim$ several hundred GeV. Collider experiments can probe higher energies ($E_{\rm lab} \sim 10^6$ GeV) but most of the information is collected in the central region. Some collider experiments such as H1 and ZEUS are able to detect forward events \cite{Abramowicz:1998ii}. The anticipated experiments LHCf~\cite{Adriani:2006jd} and TOTEM~\cite{Berardi:2004ku} in the LHC at CERN are expected to collect information in the forward region at an energy equivalent to cosmic rays of $E_{\rm lab} \sim 10^8$ GeV.

\subsection{Charged particle multiplicities}
\label{sec:nch}

Pions are the most numerous particles, followed by kaons and baryons. There is overall good agreement with experimental data in the forward region at low energies. The difference of charged particle production between the two versions is due to the improved treatment of multiple soft interactions, usage of GRV parton densities, and a consistent inclusion of diffraction dissociation, which also leads to more minijet production. These improvements give a better agreement with data for version~2.1, especially in the central region.

\begin{figure}
\includegraphics[width=0.49\textwidth]{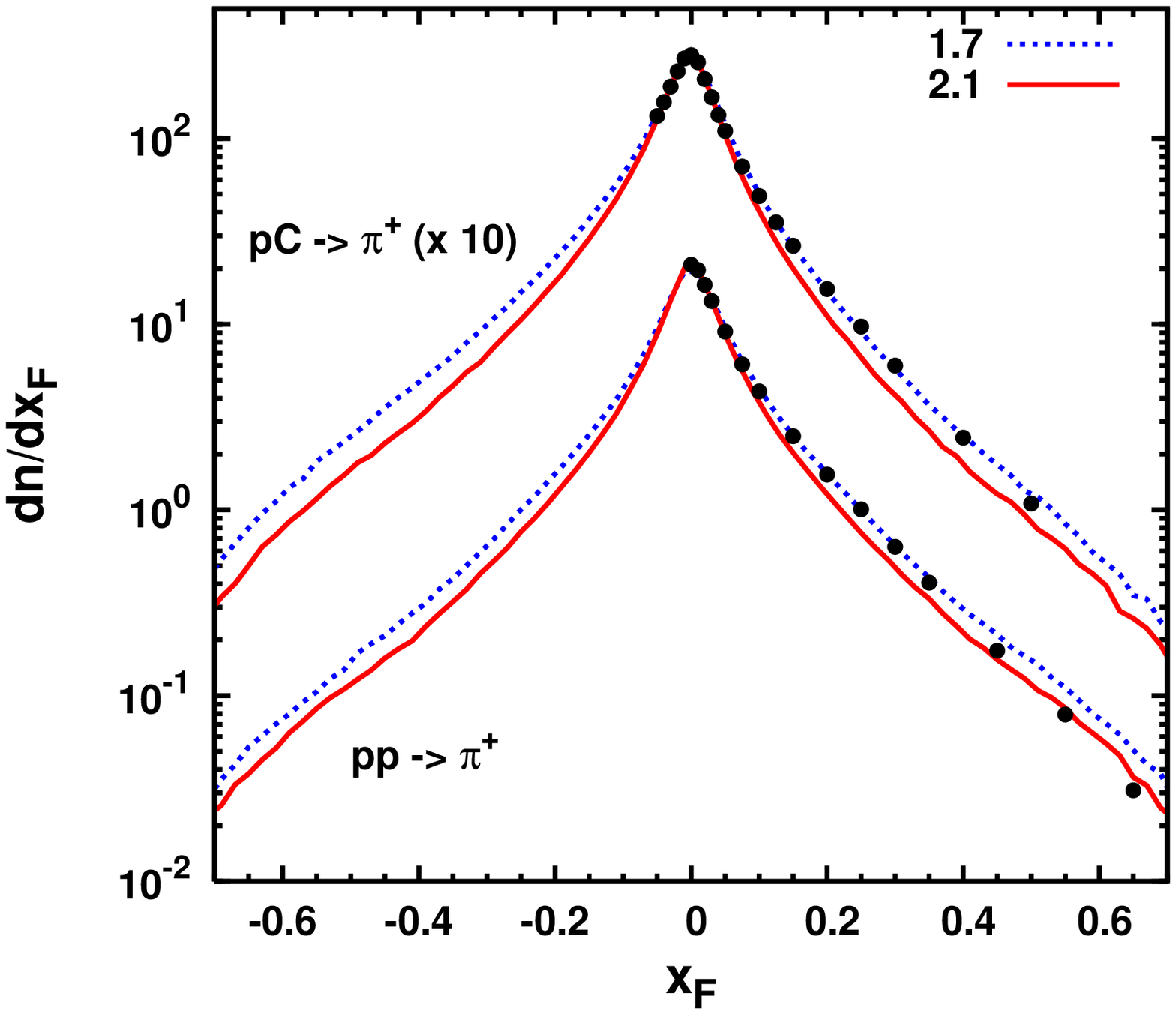}
\includegraphics[width=0.49\textwidth]{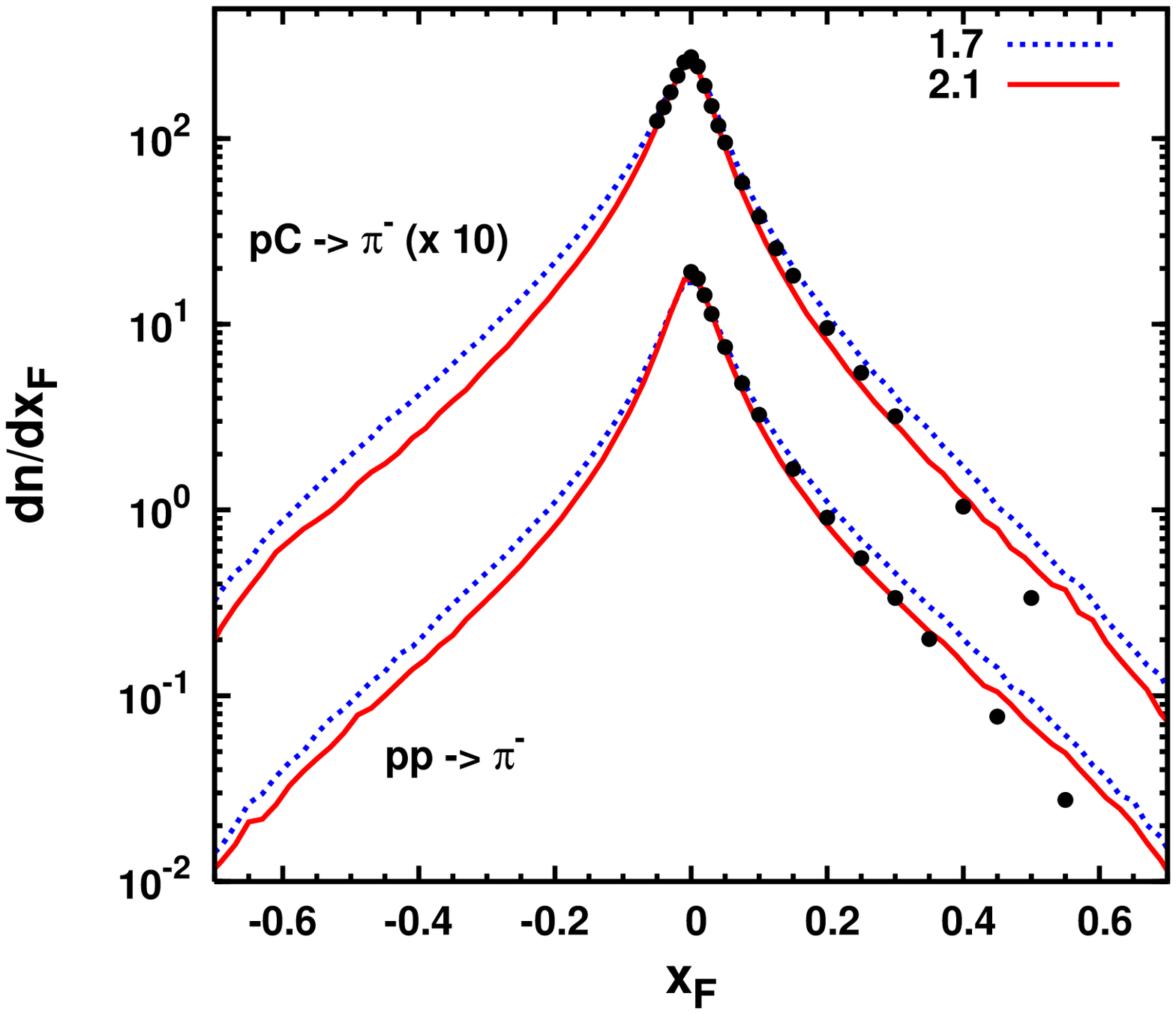}
\includegraphics[width=0.49\textwidth]{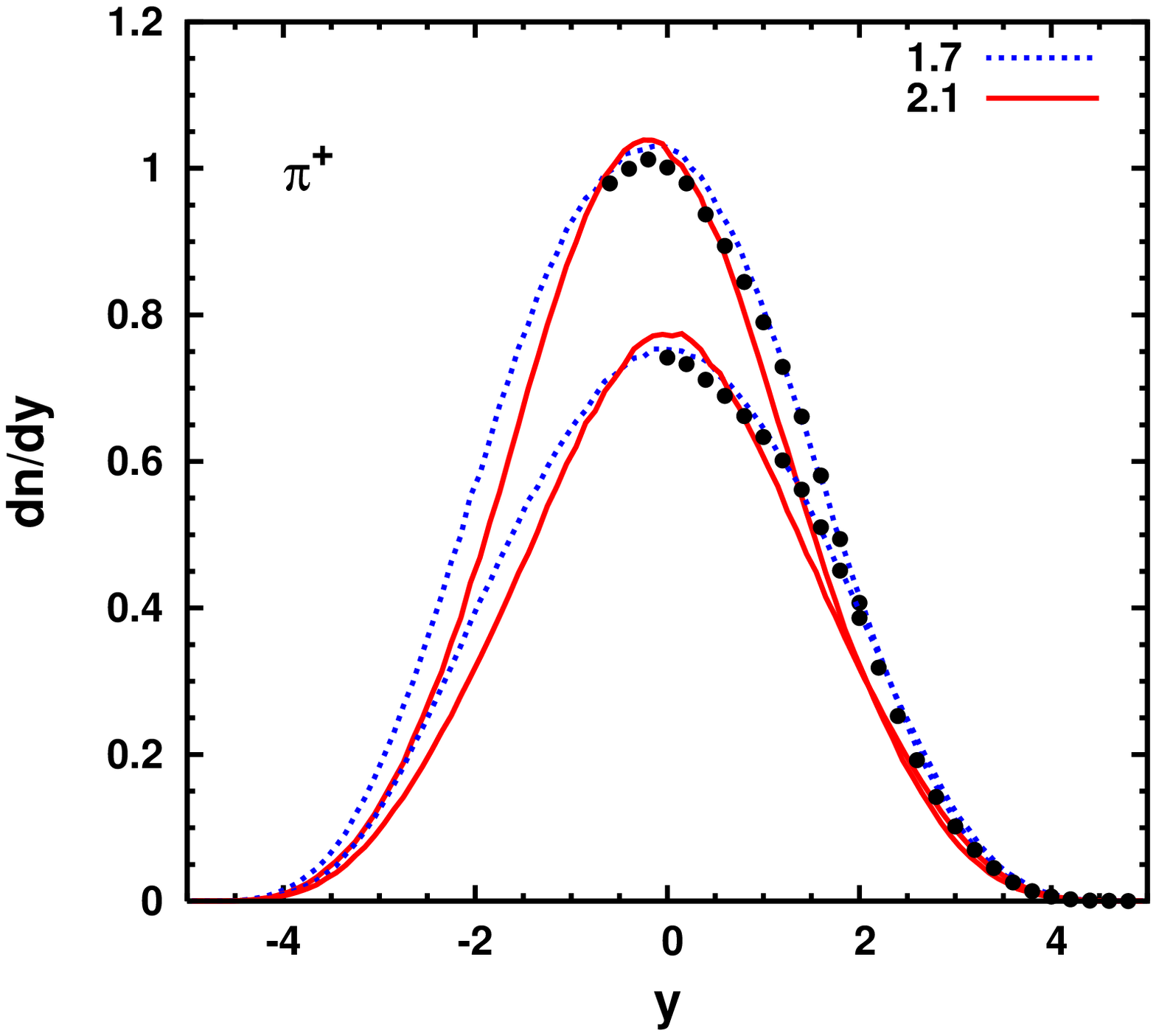}
\includegraphics[width=0.49\textwidth]{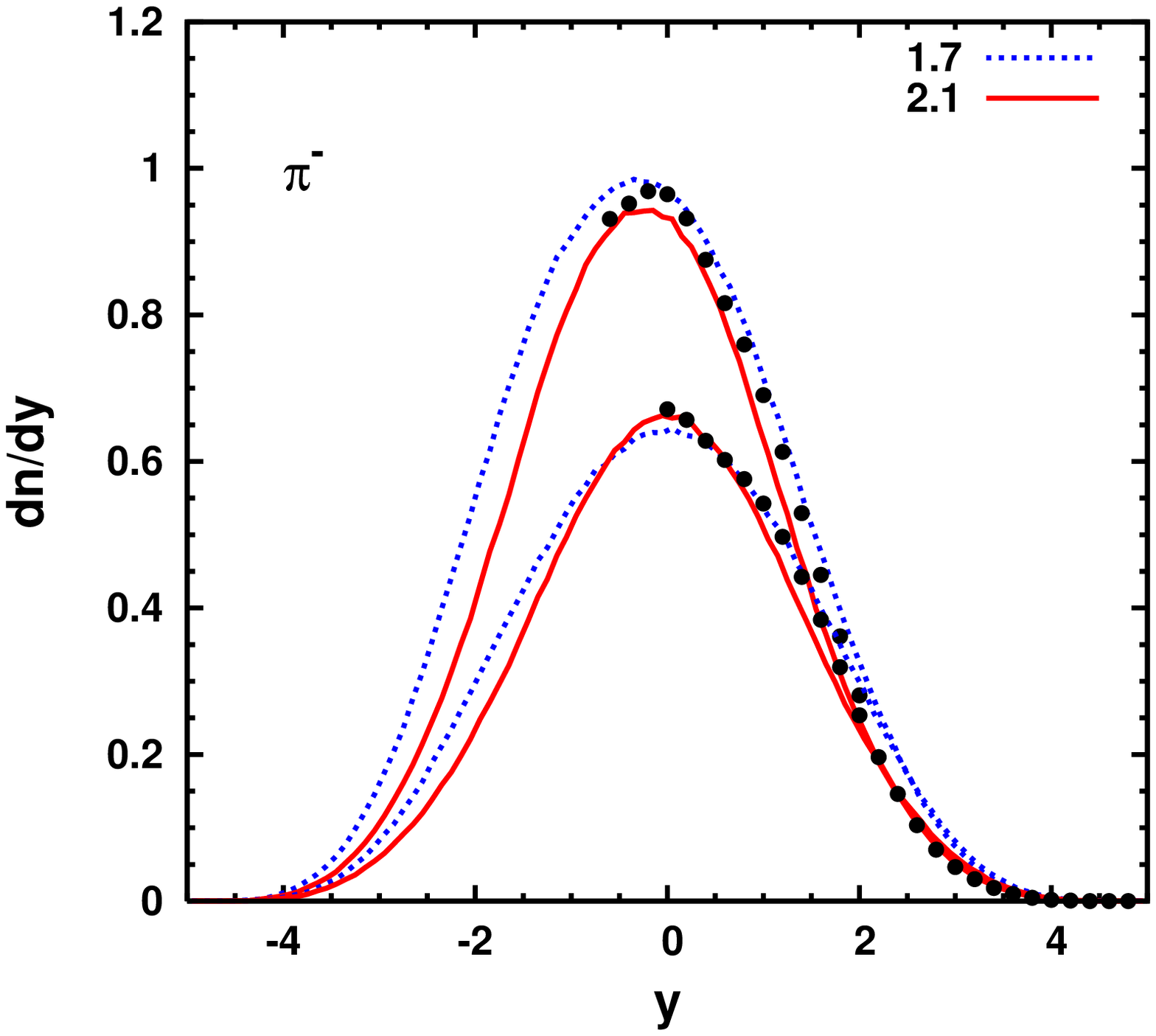}
\caption{The Feynman $x$ ($x_F$) and rapidity ($y$) distribution of pions plotted against NA49 result of $p$-$p$ \cite{Alt:2005zq} and $p$-$C$ \cite{Alt:2006fr} collision at $E_{\rm lab} = 158$ GeV. Version~2.1 (1.7) results are shown in red solid (blue dotted) lines. The left (right) panels show the production of $\pi^+$ ($\pi^-$). The upper panels show the $x_F$ distribution, where the $p$-$C$ collision results are multiplied by factor 10 in order to show both interactions on the same plot. The lower panels show the $y$ distribution: the upper (lower) set of lines and data points are from the $p$-$C$ ($p$-$p$) collision.}
\label{fig:nch-forward}
\end{figure}

The NA49 experiment measured the rapidity  $y$ and Feynman $x_F$ distribution of charged particles for $p$-$p$ \cite{Alt:2005zq} and $p$-$C$ \cite{Alt:2006fr} collisions at $E_{\rm lab} = 158$ GeV. Figure~\ref{fig:nch-forward} shows the {\sc sibyll} results compared to the data for $\pi^+$ and $\pi^-$, which became available only after the event generator had been released. Good agreement between model predictions and data is found. The excess of $\pi^+$ over $\pi^-$ is due to the flavor content of the proton ($uud$). In version~2.1, this discrepancy is stronger and more particles are produced in the central region which reflects the changes made to the soft interaction. However, the difference between the two versions is small in this respect.

\begin{figure}
\includegraphics[width=0.33\textwidth]{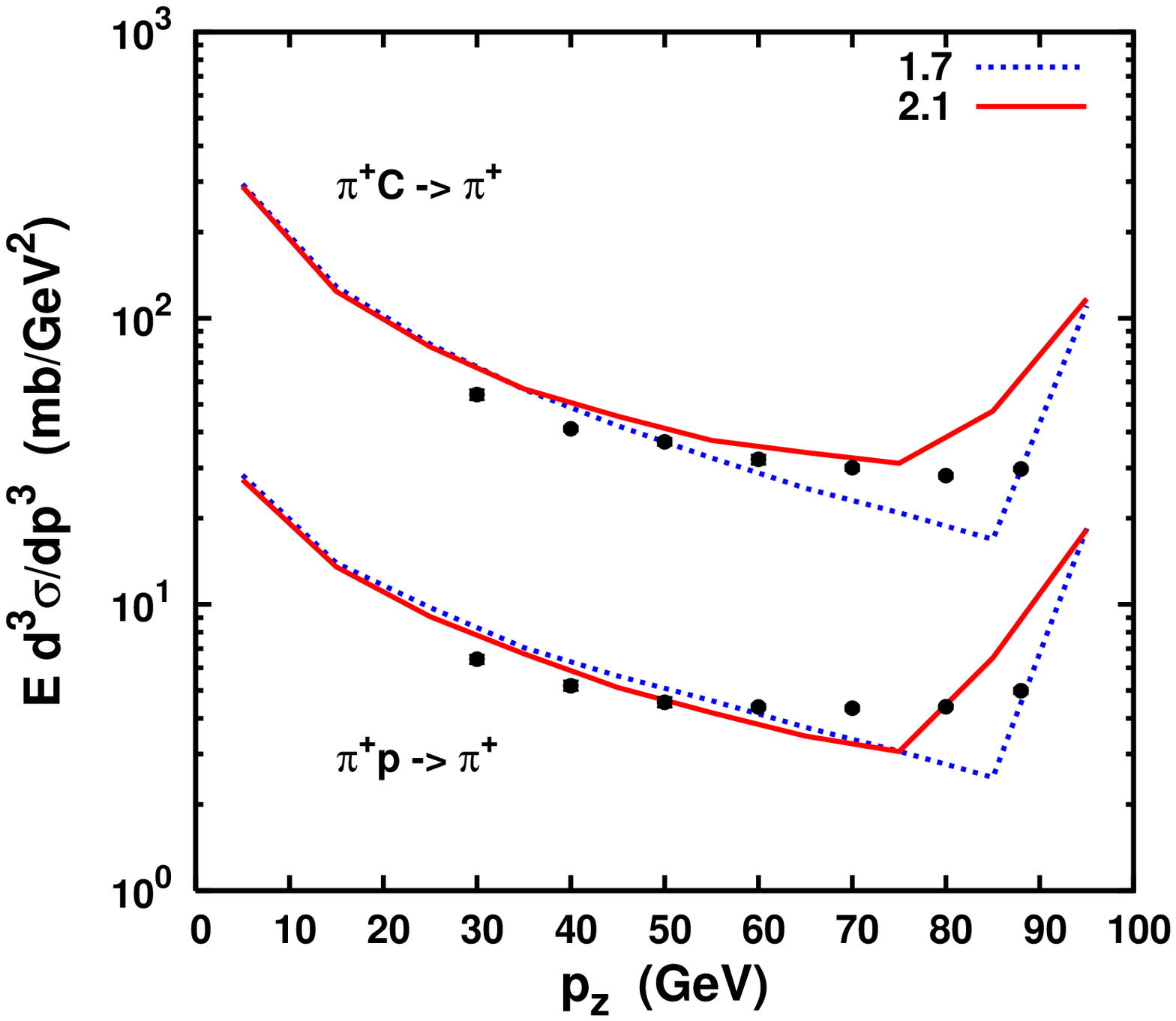}
\hspace*{-3mm}
\includegraphics[width=0.33\textwidth]{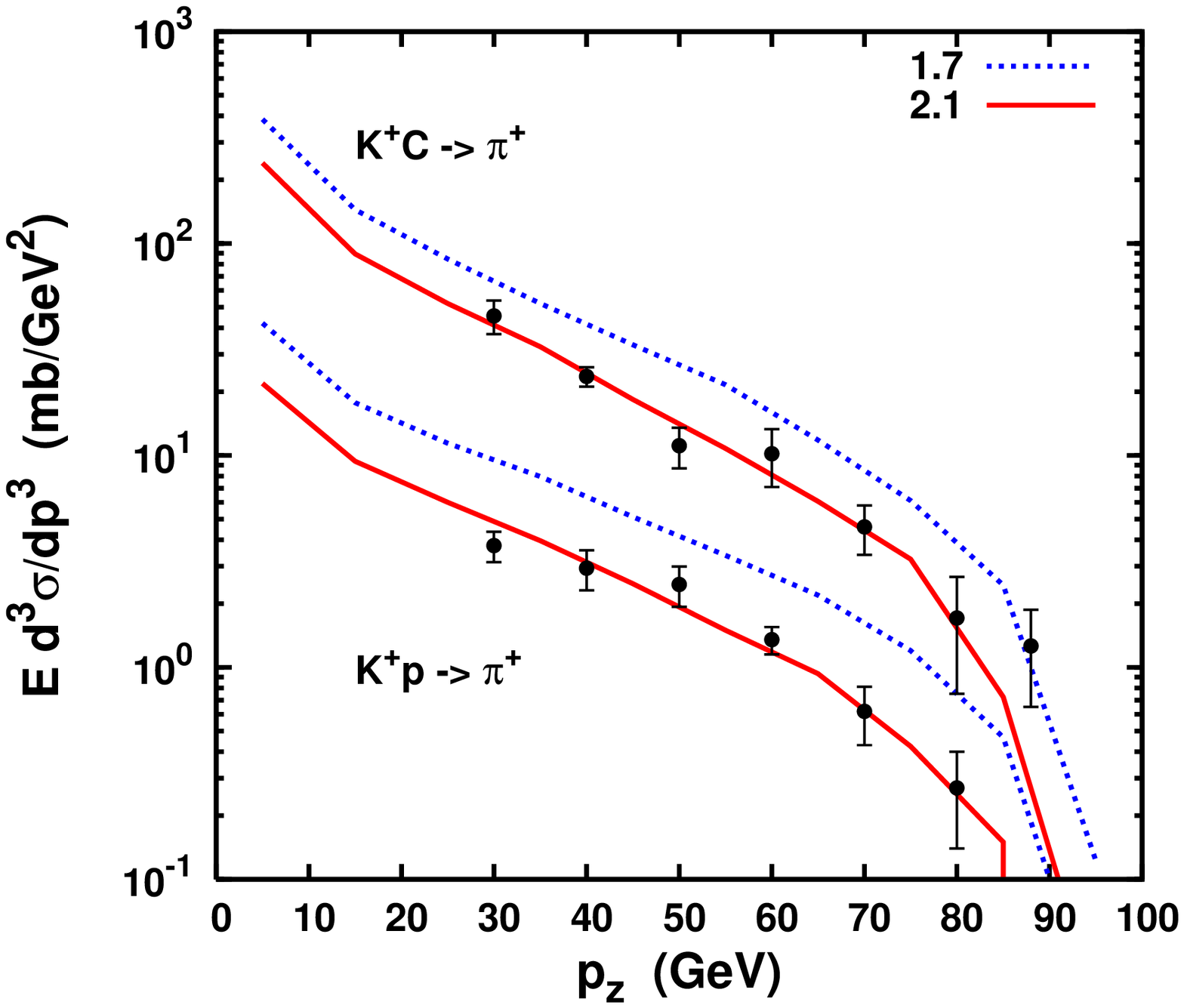}
\hspace*{-3mm}
\includegraphics[width=0.33\textwidth]{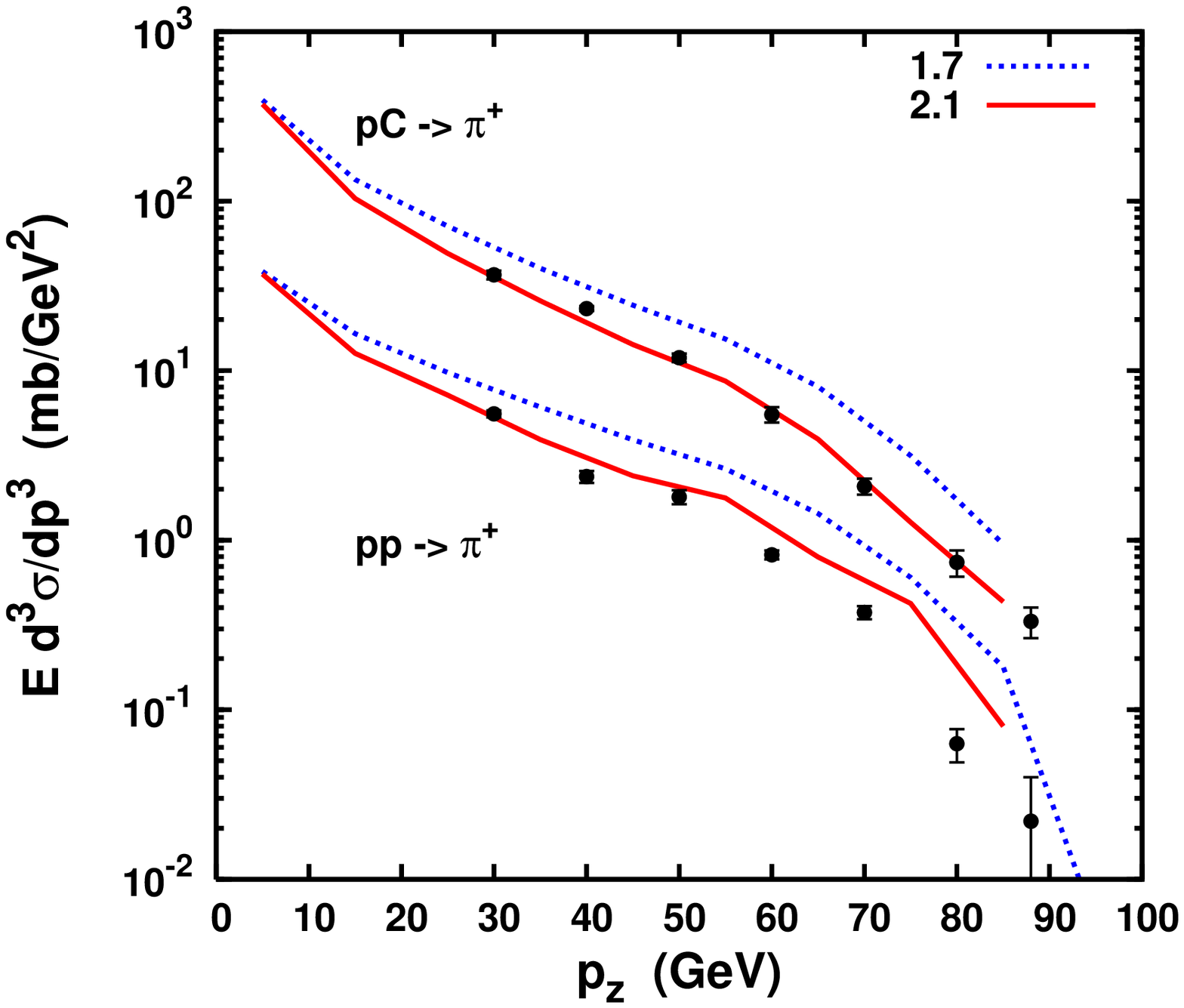}
\includegraphics[width=0.33\textwidth]{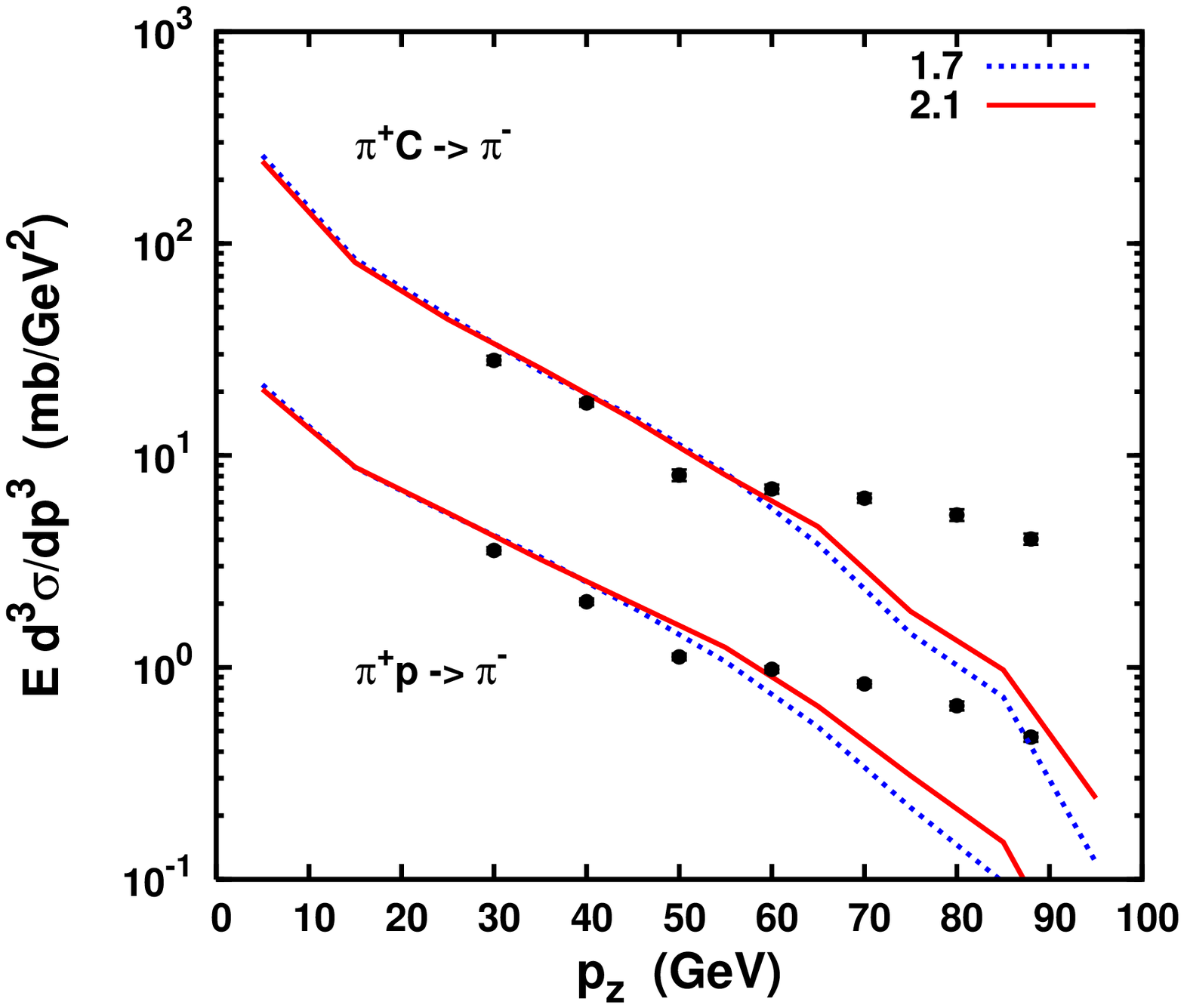}
\hspace*{-3mm}
\includegraphics[width=0.33\textwidth]{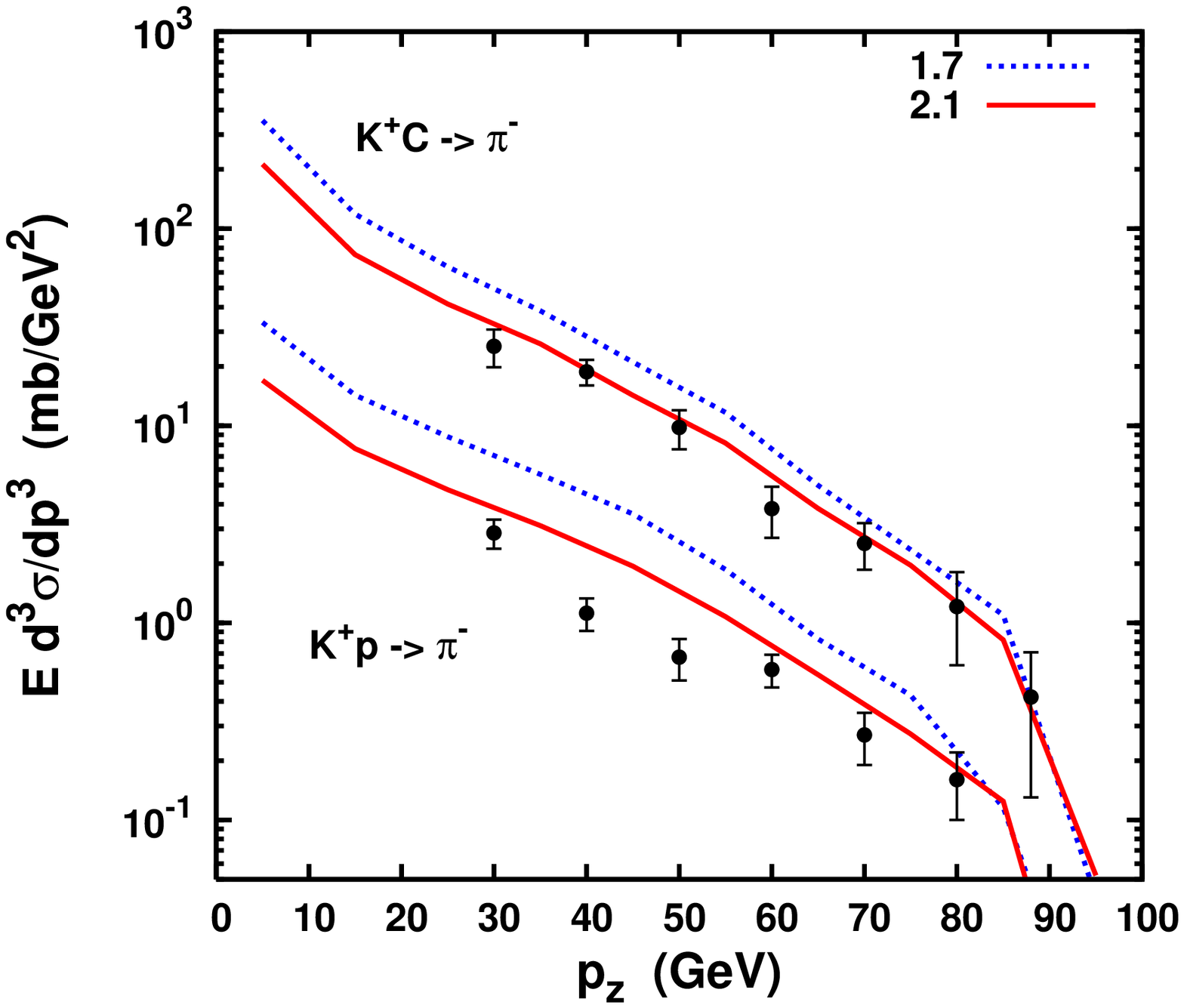}
\hspace*{-3mm}
\includegraphics[width=0.33\textwidth]{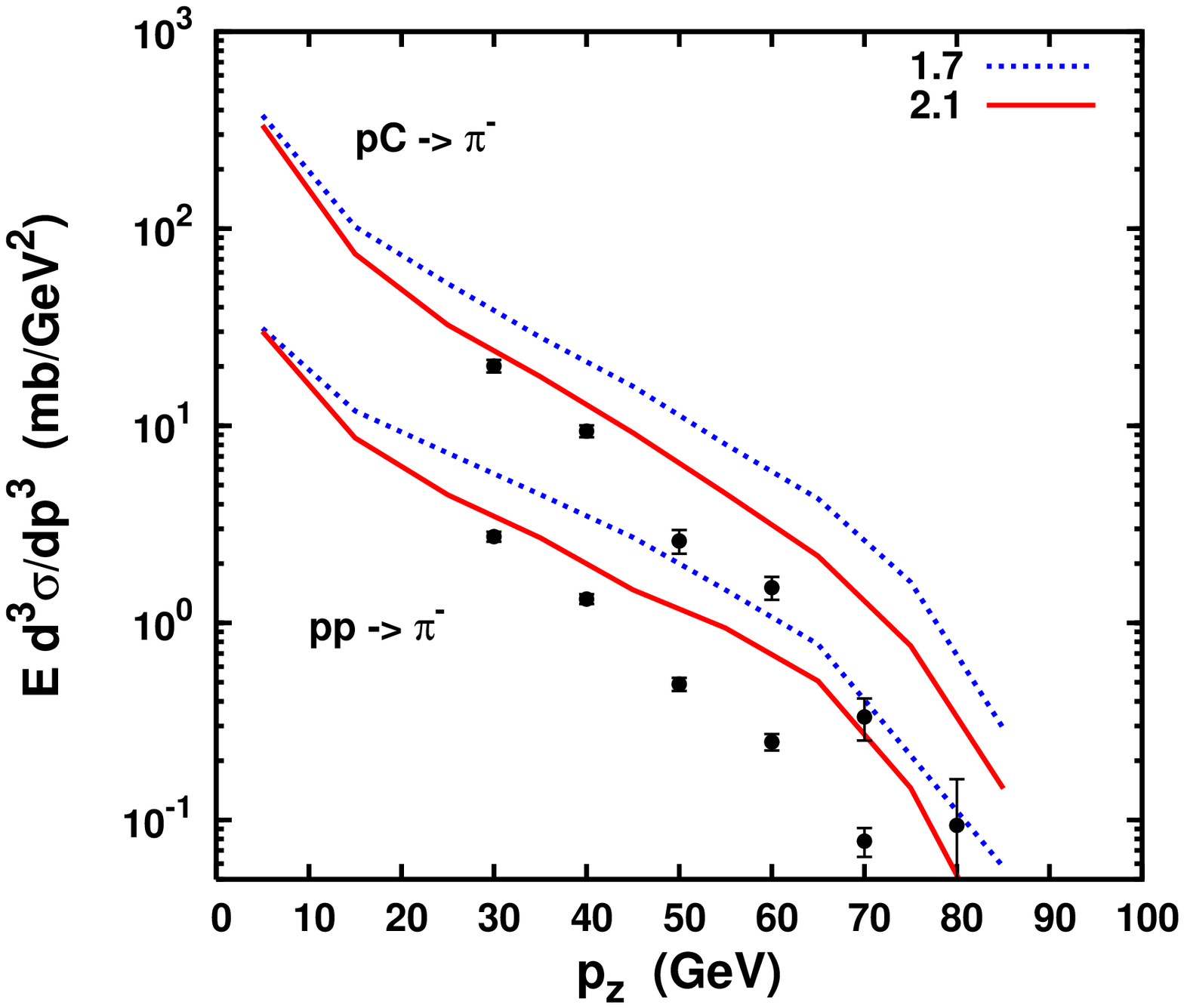}
\caption{Inclusive cross sections of $\pi^+$-$p,C$ (left panels), $K^+$-$p,C$ (center panels), $p$-$p,C$ (right panels) collisions for $\pi^+$ (upper panels) and $\pi^-$ (lower panels) production, where the events have $p_T = 0.3$ GeV. Collision energy is $E_{\rm lab} = 100$ GeV, and are compared with results from a FNAL fixed target experiment \cite{Barton:1982dg}. Version 2.1 (1.7) results are shown in red solid (blue dotted) lines.}
\label{fig:barton-pi}
\end{figure}

Fixed target experiments at FNAL used $\pi^+,~ K^+,~ p$ as projectiles and $p,~ C$ for targets. The inclusive cross section $E d^3 \sigma / d p^3$ for each charged particle species has been measured at $E_{\rm lab} = 100$ GeV at a given $p_T$~\cite{Barton:1982dg}. The results for $\pi^+ \pi^-$ production at $p_T = 0.3$ GeV/c are plotted in Fig.~\ref{fig:barton-pi}. The inclusive cross section of version 2.1 is slightly lower than version 1.7, and gives an overall better agreement. For the pion projectile, $\pi^+$s are overproduced while $\pi^-$s are underproduced in the forward region, indicating the role of the valence components in determining the leading particle. The kaon projectile shows an overall good agreement for version 2.1. For the proton projectile, there is an overproduction of $\pi^-$s compared to data, especially in the forward region. 

\begin{figure}
\includegraphics[width=0.49\textwidth]{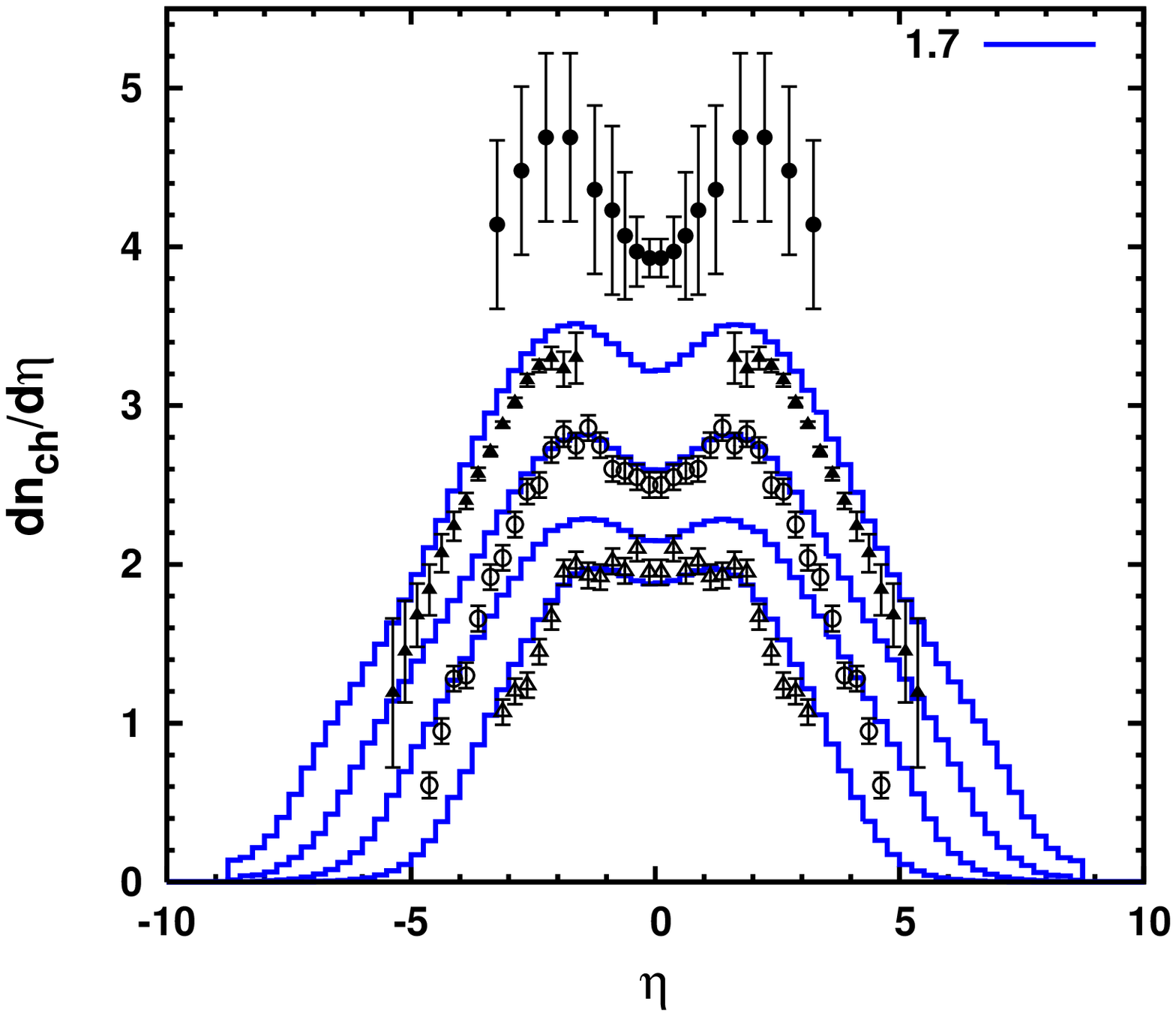}
\includegraphics[width=0.49\textwidth]{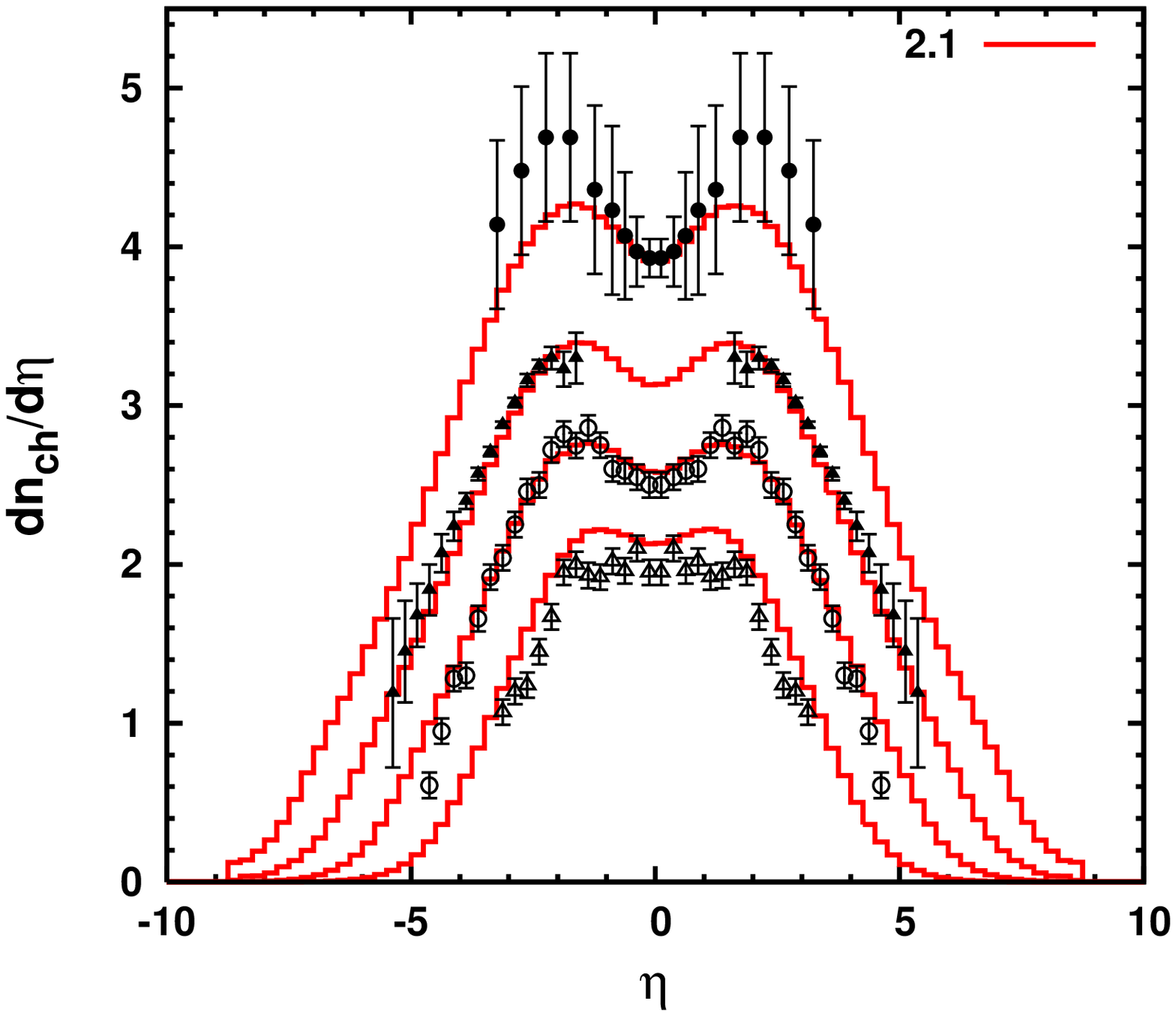}
\caption{Pseudorapidity ($\eta$) distribution of $p$-$\bar{p}$ collision, for various c.m.\ energies, for version~1.7 (left panel) and 2.1 (right panel) Data points from top are $E_{\rm c.m.} =$ 1800 GeV \cite{Abe:1989td}, 630 GeV \cite{Harr:1997sa}, 200 GeV \cite{Alner:1986xu}, 53 GeV \cite{Alpgard:1982zx}. Note the deficit of charged particles in version 1.7 at high energies in the central region.}
\label{fig:nch-pseudo}
\end{figure}

Pseudorapidity ($\eta$) distributions of charged particles from collider experiments are compared with {\sc sibyll} in Fig.~\ref{fig:nch-pseudo}.
It shows the $\eta$ distribution of charged particles from $p$-$\bar{p}$ collisions at $E_{\rm c.m.} =$ 1800 GeV (CDF~\cite{Abe:1989td}), 630 GeV (P238~\cite{Harr:1997sa}), 200 GeV (UA5~\cite{Alner:1986xu}) and 53 GeV (UA5 \cite{Alpgard:1982zx}). The improvements made to version~2.1 most prominently show in the central region. The role of the minijets and soft interactions is visible in the central region, where version~1.7 lacks secondary particles especially as the energy increases, while having more particles in the peripheral region. This trait can be seen at low energies in the $pp \to \pi^+,~ \pi^-$ figures in Fig.~\ref{fig:barton-pi}. Version~2.1 gives an excellent description of P238 data and tends to slightly overestimate the particles at low energies. It should be noted that the $\eta$ range and trigger condition for 53 GeV is different than for higher energies at UA5. The two versions are similar for events with large $|\eta|$ beyond the scope of current collider detector measurements. 

The distributions of charged particle multiplicity at UA5~\cite{Ansorge:1988kn} also give information at higher energies. 
\begin{figure}
\includegraphics[width=0.7\textwidth]{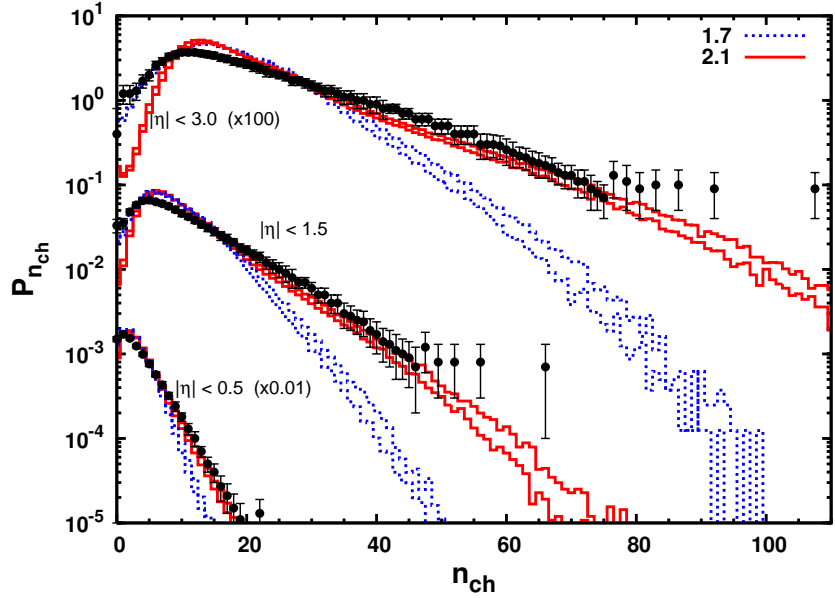}
\caption{Probability of charged particle multiplicity ($n_{ch}$) distribution at UA5 detector with $E_{\rm c.m.} = 900$ GeV for three different $\eta$ ranges; $|\eta| < 0.5$ ($\times 0.01$), $|\eta| < 1.5$, $|\eta| < 3$ ($\times 100$) \cite{Ansorge:1988kn} for version 2.1 (red solid lines) and 1.7 (blue dotted lines). The two sets of line are for having a stable and decaying $K_s^0$.}
\label{fig:nch-ua5}
\end{figure}
Figure \ref{fig:nch-ua5} shows the distribution of charged particle multiplicity for $p$-$\bar{p}$ collision at $E_{\rm c.m.} =$ 900 GeV, at three different $\eta$ ranges $|\eta| < 3.0,~ 1.5,~ 0.5$, where the results for $|\eta| < 3.0$ have been multiplied by 100 and $|\eta| < 0.5$ by 0.01 for clarity while plotting. The particle $K_s^0$ has a very short lifetime and its decay produces charged particles that have a non-negligible effect, especially at high multiplities. The treatment of $K_s^0$ is considered as one of the uncertainties in the interpretation of the experimental data. Both stable and unstable cases are plotted, which can be considered as an error band. The improvements made in soft interaction and diffraction in version 2.1 are evident in the wider distribution of $n_{ch}$ as well as in the increase in multiplicity. An underestimation of the cross section for double diffraction dissociation in {\sc sibyll} is probably the reason for the lack of low-multiplicity events satisfying the UA5 trigger condition.

\subsection{Leading particle}
\label{sec:lead}

The leading particle from the fragmentation carries a significant fraction of the total energy and becomes the primary particle in the next interaction of the air shower. The elasticity $K = E_{\rm lead}/E_{\rm proj}$, the fraction of the leading particle with respect to the collision energy, of a collision affects the multiplicity as well as the speed of shower development in the atmosphere. Thus it is important to get a correct description of the behavior of the leading particles.

\begin{figure}
\includegraphics[width=0.6\textwidth]{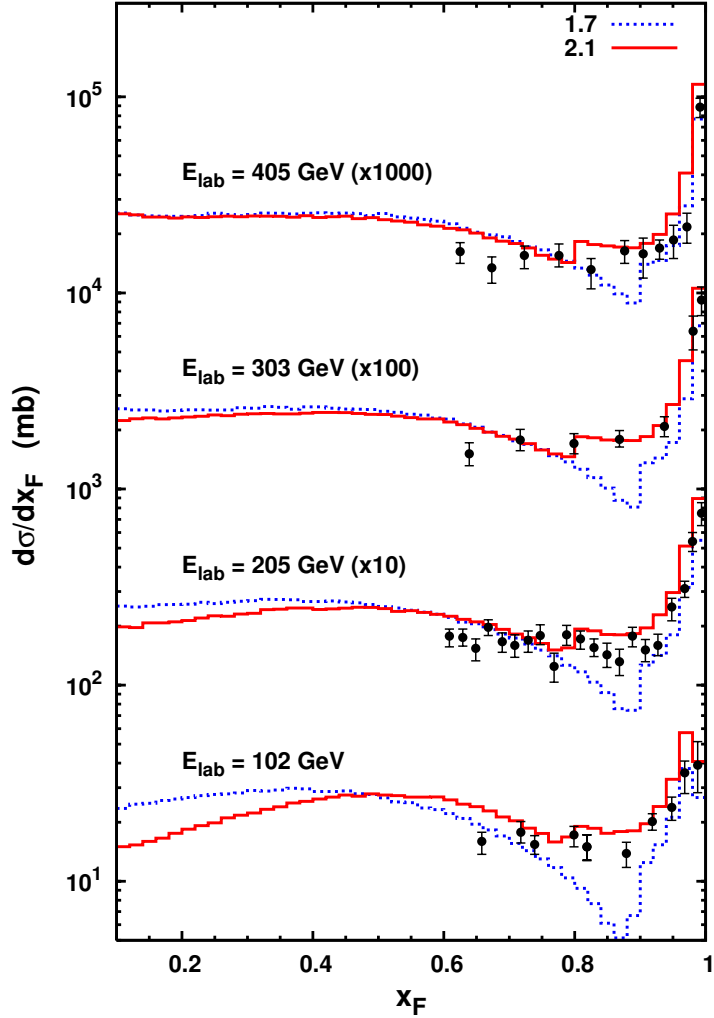}
\caption{Distribution of leading protons as a function of $x_F$ compared to the $p$-$p$ collision data from the NAL bubble chamber \cite{Whitmore:1973ri}. Version~2.1 (1.7) results are shown in red solid (blue dotted) lines. The results and datapoints have been multiplied by factor of tens (405 GeV by 1000, 303 GeV by 100, 205 GeV by 10, 102 GeV by 1).}
\label{fig:nal-lead}
\end{figure}
The NAL bubble chamber experiment has data for $p$-$p$ interactions at $E_{\rm lab} = $ 102, 205, 303, 405 GeV and measured the $x_F$ of the leading proton~\cite{Whitmore:1973ri}. Figure~\ref{fig:nal-lead} shows the {\sc sibyll} results plotted against the NAL data. The sharp dip at $x_F \approx 0.9$ for the old version indicates the abrupt onset of diffraction, which is softened for version~2.1. It is not a smooth turn-on however, with the switch-on evident from the small step around $x_F = 0.8$. 

\begin{figure}
\includegraphics[width=0.6\textwidth]{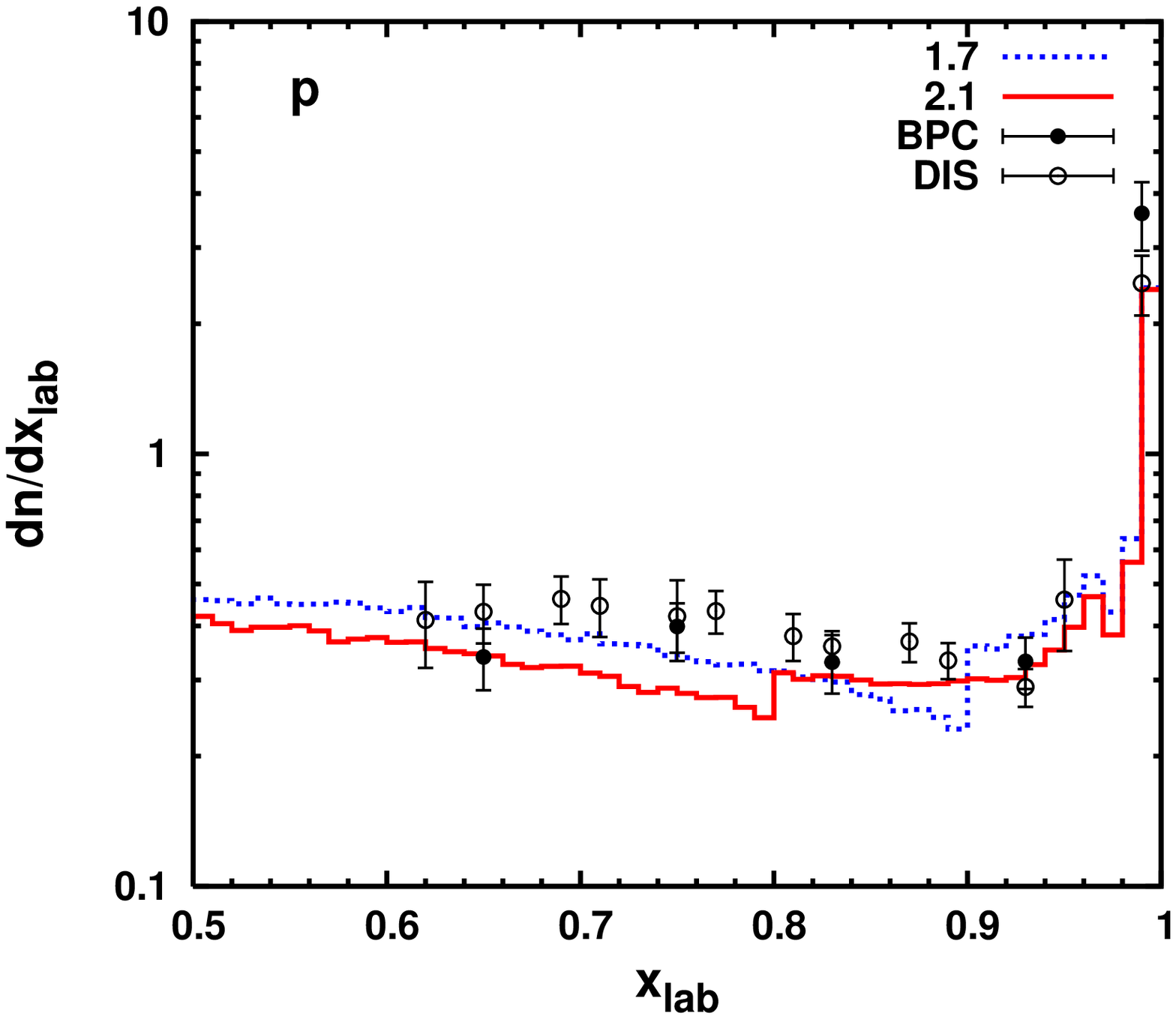}
\caption{Leading proton distribution of {\sc sibyll} compared to the low-$Q^2$ [beam pipe calorimeter (BPC)] and higher-$Q^2$ [deep inelastic scattering (DIS)] ZEUS data \cite{Chekanov:2002yh, Chekanov:2002pf}. Version~2.1 (1.7) outcome is shown in red solid (blue dotted) lines. The transverse momentum of the protons are limited to $p_T^2 < 0.5$ GeV$^2$. The c.m.\ collision energy of {\sc sibyll} is 210 GeV.}
\label{fig:zeus-lead}
\end{figure}
The ZEUS detector measured the leading proton~\cite{Chekanov:2002yh} with small transverse momentum. ZEUS collided positrons with protons, where the proton energy was 820 GeV during run I. The c.m.\ collision energy is about 300 GeV. A virtual photon emitted from the positron interacts with the proton. The ZEUS Collaboration has confirmed that the initial projectile gives little effect~\cite{Chekanov:2002pf} on the outcome. As {\sc sibyll} cannot have a photon or positron projectile,
 we simulated a $p$-$p$ collision at a slightly lower energy of $E_{\rm c.m.} = 210$ GeV and used events from one hemisphere, i.e. events with $p^{\rm c.m.}_z > 0$. Figure~\ref{fig:zeus-lead} shows the leading protons of the {\sc sibyll} results plotted against the ZEUS data. They are plotted as a function of $x_{\rm lab} = E / E_p$, the energy of the proton or neutron divided by the colliding proton energy in the lab frame, which is essentially the elasticity. The leading proton displays similar behavior to that of the NAL bubble chamber. Again, the better diffraction treatment is evident around $x_{\rm lab} = 0.9$. 

\subsection{Strange particle production}

\begin{figure}
\includegraphics[width=0.49\textwidth]{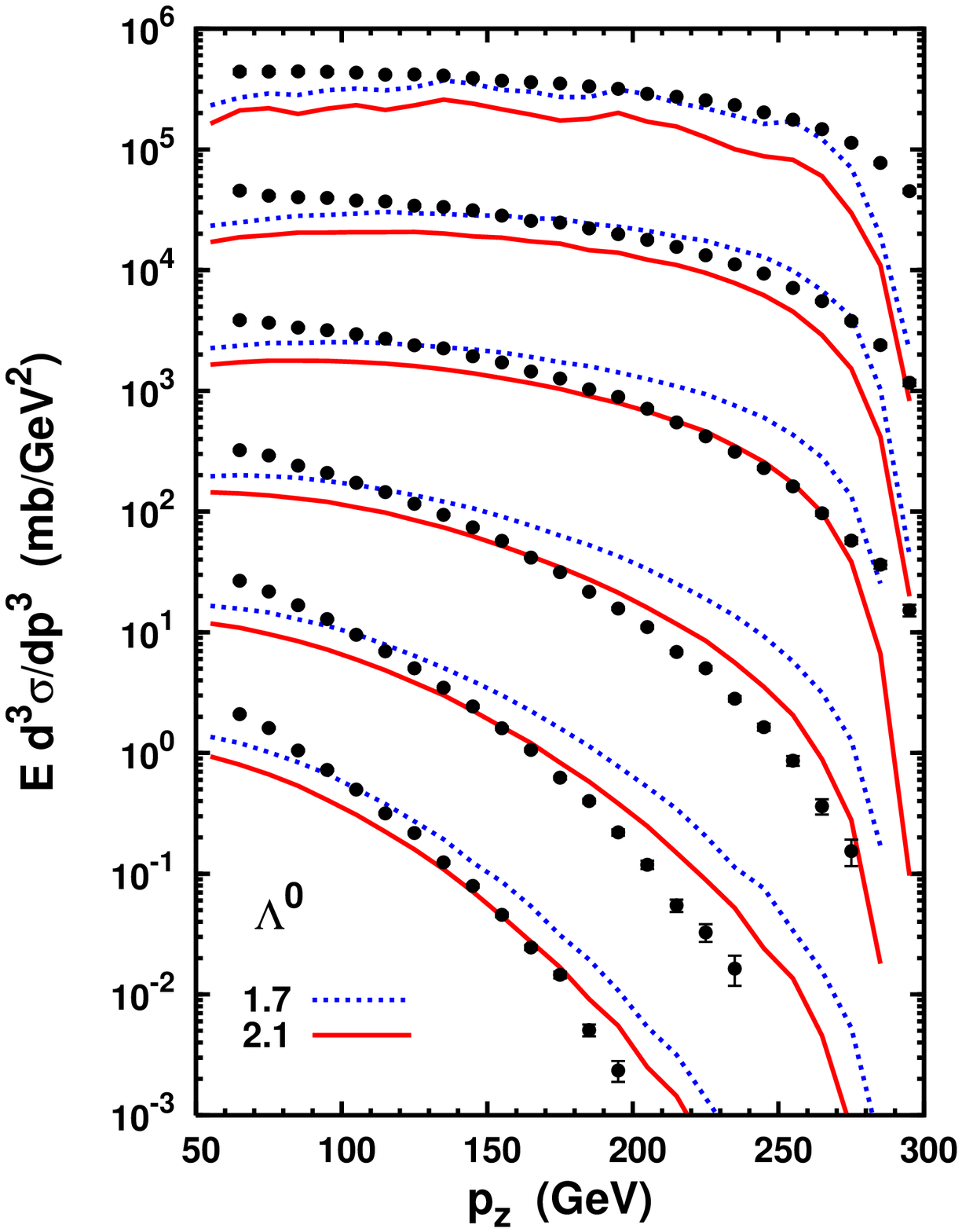}
\includegraphics[width=0.49\textwidth]{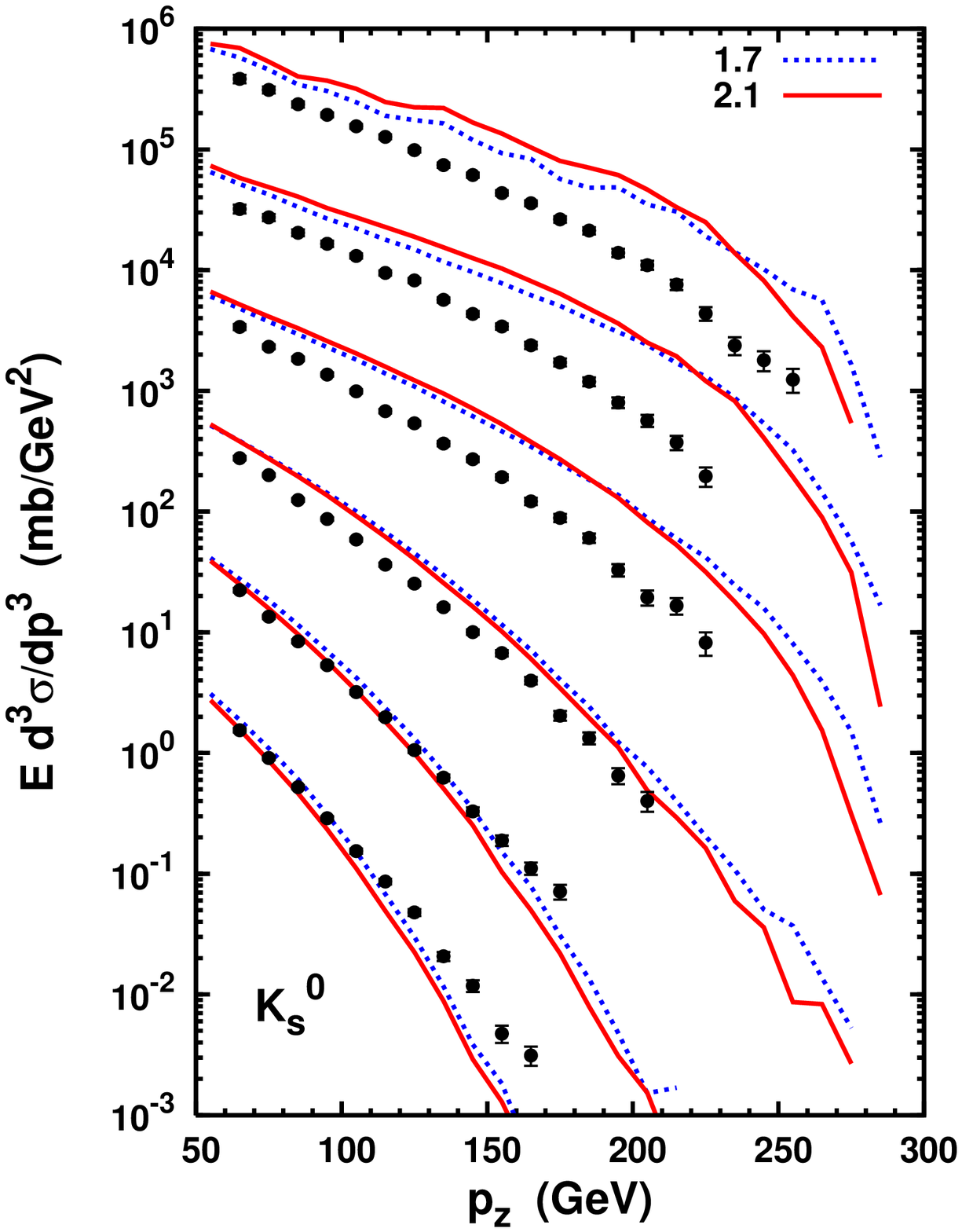}
\caption{Inclusive cross sections of $\Lambda^0$ (left panel) and $K_s^0$ (right panel) production from $p$-$Be$ collision at $E_{\rm lab} = 300$ GeV for various angles \cite{Skubic:1978fi}. Version~2.1 (1.7) results are  shown in red solid (blue dotted) lines. From the top, the sets of lines and data are for angles 0.25, 1.5, 2.9, 5.0, 6.9, 8.8 mrad. The results and data have been multiplied by factors of tens to show them in one figure (0.25 mrad by $10^5$, 1.5 mrad by $10^4$, 2.9 mrad by $10^3$, 5.0 mrad by 100, 6.9 mrad by 10, 8.8 mrad by 1).}
\label{fig:strange}
\end{figure}

A FNAL fixed target experiment measured the production of very forward strange particles produced in $p$-$Be$ collisions at $E_{\rm lab} = 300$ GeV~\cite{Skubic:1978fi}. The inclusive cross section of $\Lambda^0$ and $K_s^0$ have been plotted for angles in the range $\theta = 0.25 - 8.8$ mrad in Fig.~\ref{fig:strange}.
As these are very forward direction measurements, particles with large $p_z$ are likely to be from particle production associated with the projectile, and particles with small $p_z$ are from central production. Both {\sc sibyll} versions give agreements in the forward direction, with a tendency to slightly overproduce high-$p_z$ particles and underproduce low-$p_z$ particles. Strange particle production directly affects production of high energy muons and neutrinos.

\section{Air shower performance}

The development of an air shower depends on a number of factors, some of which are the production cross section, inelasticity, and multiplicity. For the description of the air shower development the hadronic interaction model has to describe correctly the  particle interactions in a wide range of energies. Observables such as depth of shower maximum $X_{\rm max}$, electron number $N_e$ and muon number $N_\mu$ at ground will depend on the characteristics of hadronic interactions\footnote{Extensive studies have been carried out in Ref.~\cite{Ulrich:2009hm}, where cross sections, elasticity, and multiplicity are varied to see the effect on observables such as $X_{max}$, $N_e$ and $N_\mu$.}. We briefly summarize how air showers are affected by those three parameters.

Increasing the cross section will cause the shower to start earlier in the atmosphere, resulting in a smaller $X_{\rm max}$ as well as a smaller fluctuation. The number of electrons measured is highly dependent on the position of the shower maximum; the closer to $X_{\rm max}$ the larger the $N_e$. As most muons are produced from decay of pions and kaons, $N_\mu$ is expected to remain stable. 

An increase in the mean multiplicity lessens the energy per particle of the secondaries, which results in a quicker development of the shower with smaller fluctuations. The increased multiplicity is expected to increase $N_\mu$. The number of electrons are most numerous at $X_{\rm max}$ and decreases away from it. Hence, a quicker development results in a larger distance between $X_{\rm max}$ and ground which will contribute to decreasing the number of electrons detected on ground.

A large elasticity gives a larger fraction of the total energy to the leading particle, and slows the shower development as well as giving a smaller multiplicity. The $X_{\rm max}$ will be closer to the ground and $N_\mu$ will be smaller. The closer proximity of $X_{\rm max}$ to the ground causes a larger $N_e$ despite the smaller multiplicity~\cite{Ulrich:2009hm}.

\begin{figure}
\includegraphics[width=0.65\textwidth]{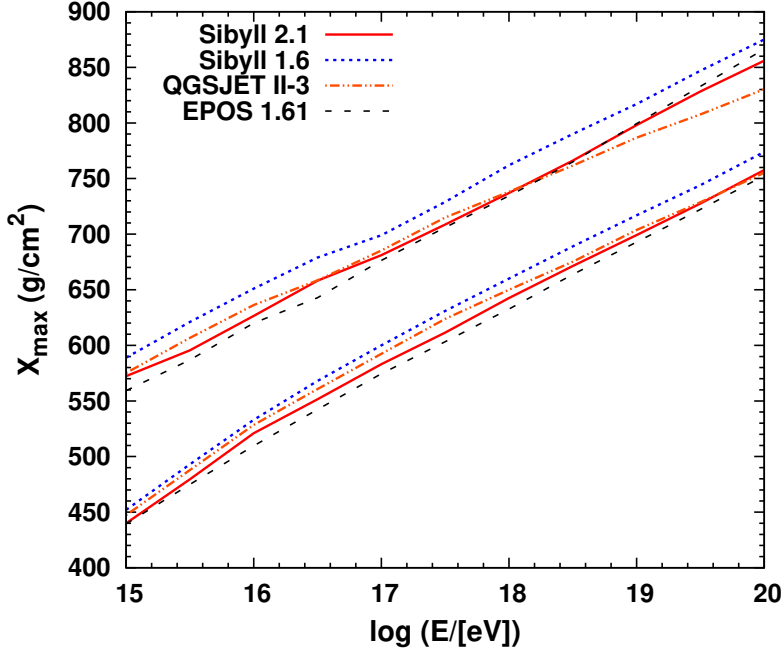}
\caption{The mean $X_{\rm max}$ of protons and iron nuclei as a function of primary energy at zenith angle $\theta = 45^\circ$. In the case of {\sc sibyll} (red solid lines), {\sc qgsjet}~II-3 (orange dot-dashed lines), and {\sc epos}~1.61 (black dashed lines), the data are taken from the hybrid simulator CONEX \cite{Pierog:2004re, Bergmann:2006yz} with statistical uncertainly of typically $1 \!-\! 3~ {\rm g\,cm}^{-2}$. For {\sc sibyll} 1.6 (blue dotted lines), 500 events were simulated with Corsika 5.62 \cite{Heck:1998vt} (from Ref.~\cite{Pryke:2000ac}) with statistical uncertainty of $1 \!-\! 4~ {\rm g\,cm}^{-2}$.
\label{fig:xmax}}
\end{figure}
The new version has the shower developing more quickly than the old version. The two versions are compared with {\sc qgsjet}~II-3~\cite{Ostapchenko:2005nj} and {\sc epos}~1.6~\cite{Pierog:2006qv}. The shower maximum is shown in Fig.~\ref{fig:xmax}. {\sc sibyll}~2.1 has an increased cross section and larger multiplicity which results in a smaller $X_{\rm max}$ overall but maintains the same shape. {\sc qgsjet}~II-3 has a very large multiplicity at high energies, resulting in smaller shower maximum.  

\begin{figure}
\includegraphics[width=0.75\textwidth]{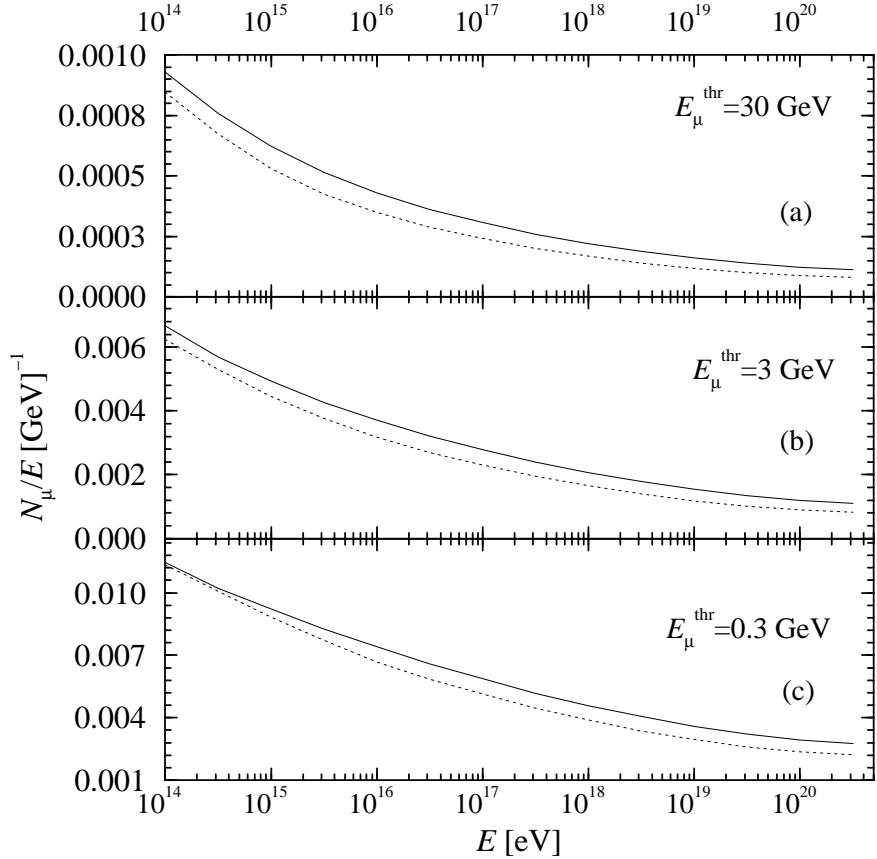}
\caption{Average number of muons at sea level $\langle N_\mu\rangle$, obtained in proton showers with zenith angle $\theta=0^{\circ}$. Each energy represents 5000 showers simulated with the hybrid method. The solid (dotted) line represents the values obtained with {\sc sibyll} 2.1 ({\sc sibyll} 1.7). Panels (a)-(c) show the average number of muons with energy above 30, 3 and 0.3 GeV respectively. Figure is adapted from Fig.~14 of Ref.~\cite{Alvarez-Muniz:2002ne}.}
\label{NmuModelsPlot}
\end{figure}

The muon number is an important indicator for cosmic ray composition studies, as showers of heavier nuclei contain more muons than that of protons. Version~2.1 produces more muons than version 1.7, which cause the new version of {\sc sibyll} to extract a lighter cosmic ray composition from experimental data than the old one. However, both versions produce fewer low energy muons than other models such as {\sc qgsjet}~II-3 or {\sc epos}~1.6. The muon number at sea level is sensitive to zenith angle. When the atmosphere depth increases, muon production and muon energy loss and decay compete against each other. For low muon energies, the decay process is dominant and $N_\mu$ decreases as increasing zenith angle. At sufficiently high energies above $\sim 10$ GeV, most muons do not decay and $N_\mu$ increases. Figure~\ref{NmuModelsPlot} shows the energy dependence of the average number of muons normalized to the primary energy for the two versions of {\sc sibyll}. The average number of muons at sea level ($\langle N_\mu\rangle$) is plotted with energies above $E_\mu^{\rm thr}=$0.3, 1, 3, 10 and 30 GeV in proton-initiated showers at zenith angle $\theta=0^\circ$.

\section{Shortcomings and future improvements}

Though many features have been improved, {\sc sibyll} 2.1 is by no means a complete model. We list the shortcomings of the current version and describe how we plan to improve the model.

\begin{itemize}

\item The current nucleus-nucleus collision uses the semisuperposition model. Implementing the full  Glauber model will give a more accurate description.

\item Antibaryon production is not satisfactorily described. There is not enough produced and the distribution of antiprotons in the central region is incorrect compared with data~\cite{Alexopoulos:1993wt}. The overall normalization can be improved by increasing the diquark production fraction. The current method of fragmentation suppresses antibaryon formation together with other particles in the non-end of strings. 

\item The currently used energy-dependent transverse momentum cutoff is independent of the relevant gluon density of the interaction with the target nucleus, which varies with the impact parameter of the collision. A new energy- and impact parameter-dependent cutoff to $p_T$ to prevent parton density saturation would improve the modeling. Reference~\cite{Rogers:2008ua} is an example of an attempt to constrain minijet formation. In addition the profile functions could also be more refined.

\item A consistent treatment of coherent and incoherent diffraction dissociation in hadron-nucleus and nucleus-nucleus interactions is required. This can be achieved by using a two-channel model in the Glauber calculation similar to the one presented here for $p$-$p$ interactions.

\item Include charm quark. The current model has $u,~d~s$ quarks and gluons. This improvement will be relevant more to muon and neutrino detectors than cosmic ray observations. A version containing charm will be released soon.

\end{itemize}

\section{Conclusion}

In this paper, we described the overall model of {\sc sibyll} together with the changes made in the version 2.1, and listed the shortcomings and possible ways of improving the model. The 2.1 version still keeps the DPM-minijet structure but with modifications and additions. Results from HERA suggest a steeper parton density for gluons at low $x$, and the GRV parton densities replaced the EHLQ parton densities which resulted in an increased QCD cross section. Concepts from Regge theory are used to allow multiple soft interactions. The energy-dependent transverse momentum cutoff for ensuring perturbative QCD to be valid is a better discriminator between soft and hard interactions than the previously applied energy-independent cutoff. The two-channel eikonal model is used to describe diffraction, with better if not complete success. The physics framework for the hadron-nucleus and nucleus-nucleus interactions remains unchanged.

These improvements produce more particles with a wider distribution in momentum space, as are evident when comparing with experimental data. Both versions give a good fit to the rapidity and Feynman $x$ distribution for fixed target experiments such as NA49. However the changes made to the new version are very evident in the central region, in the pseudorapidity and the overall charged particle multiplicity distribution. As a consequence, air showers described with {\sc sibyll}~2.1 develop quicker, with smaller shower maximum and larger muon number than version 1.7.

{\sc sibyll}~2.1 is by no means the final answer. There are still shortcomings, such as unsatisfactory description of antibaryon production, and further room for improvements in preventing parton density saturation by giving an energy-dependent cutoff to the transverse momentum, giving a better diffraction treatment and have a full Glauber model for nucleus-nucleus interaction. A version including charm quark production is due to be released shortly.

With all its advantages and shortcomings, the interaction model {\sc sibyll}~2.1 is able to successfully reproduce and predict many experimental data. We believe there that use of {\sc sibyll} in comparison with other models would be valuable in analysis of cosmic ray air shower data.

\acknowledgments

Work on this project at the University of Delaware is supported in part by a grant from the Office of Science of the U.S. Department of Energy, DE-FG02-91ER40626. Fermilab is operated by Fermi Research Alliance, LLC under Contract No. DE-AC02-07CH11359 with the United States Department of Energy.

\bibliographystyle{h-physrev3}
\bibliography{sibyll-21.bib}

\appendix

\section{Amplitude conventions}
\label{app:conv}

The conventions and parameters are adopted from Ref.~\cite{Block:1984ru}. The Mandelstam variables $s$, $t$, $u$ are used, and $k$ is the c.m.\ momentum. The c.m.\ scattering angle $\theta$ is related to $t$ by
\be
t \,=\, - 4 k^2 \sin^2(\theta/2) \ ,
\label{eq:mandel-t}
\ee
and a new parameter is defined $-q^2 \equiv t$. The scattering amplitude in the c.m.\ frame [$f_{\rm c.m.}(s,t)$] and the Lorentz invariant scattering amplitude [$\mathscr{M}(s,t)$] are related by
\be
\mathscr{M}(s,t) \,=\, -8 \pi \, \sqrt{s} \, f_{\rm c.m.}(s,t) \ .
\ee

The scattering amplitude $f_{\rm c.m.}(s,t)$ and impact parameter function
$a(s,\b)$ are Fourier transforms of each other via
\ba
\displaystyle
f_{\rm c.m.}(s,t) &=& \frac{k}{\pi} \int d^2 b \, e^{i {\bf{q}} \cdot \b} \, a(s,\b) \ , \label{eq:fcm} \\
a(s,\b) &=& \frac{1}{4 \pi k} \int d^2 q \, e^{-i {\bf{q}} \cdot \b} \, f_{\rm c.m.}(s,t) \ . 
\ea
With this scattering amplitude, the differential elastic cross section can be expressed as
\begin{subequations}
\ba
\frac{d \sigma_{\rm el}}{d \Omega_{\rm c.m.}} &=& | f_{\rm c.m.}|^2 \ , \label{eq:app-dsigdom} \\
\frac{d \sigma_{\rm el}}{d t} &=& \frac{\pi}{k^2} \, | f_{\rm c.m.}|^2 \ , \label{eq:app-dsigdt} \\
\sigma_{\rm tot} &=& \frac{4 \, \pi}{k} \, {\rm Im} f_{\rm c.m.}(\theta = 0) \ ,
\ea
\end{subequations}
where the last relation used the optical theorem. The elastic slope parameter $B_{\rm el}$ is defined from an approximation of the elastic scattering cross section in small $t$ region as
\be
\frac{d \sigma_{\rm el}}{dt} ~=~ \left[ \frac{d \sigma_{\rm el}}{dt} \right]_{t=0} \, e^{B t} \ .
\ee
Using Eqs.~\refb{eq:app-dsigdom} and \refb{eq:app-dsigdt},
\be
\displaystyle
\begin{split}
\left[ \frac{d \sigma_{\rm el}}{dt} \right]_{t=0} 
&=~ \frac{\pi}{k^2} \, \left[ \frac{d \sigma_{\rm el}}{d \Omega_{\rm c.m.}} \right]_{\theta=0} \\
&=~ \pi \, \left| \frac{ (\rho \,+\, i) \, {\rm Im} f_{\rm c.m.}(0)}{k} \right|^2 
~=~ \pi \, \left| \frac{ (\rho \,+\, i) \, \sigma_{\rm tot}}{4 \, \pi} \right|^2 \ ,
\end{split}
\ee
where $\rho$ is the real to imaginary ratio of $f_{\rm c.m.}(0)$.

When using eikonals, the cross sections follow the usual convention:
\ba
\displaystyle
\sigma_{\rm tot} ~=~ 2 \pi \int d b^2 (1 \,-\, e^{-\chi(b,s)}) \\
\sigma_{\rm el} ~=~ \pi \int d b^2 (1 \,-\, e^{-\chi(b,s)})^2 \\
\sigma_{\rm inel} ~=~ \pi \int d b^2 (1 \,-\, e^{-2\chi(b,s)})\ .
\ea
We neglect the real part of the elastic scattering amplitude for calculating the eikonal functions.

\section{Hard interaction minijet cross section and profile functions.}
\label{app:hard}

Each proton or meson is characterized by a (transverse) density profile function $A_z(\b)$, where $z$ can be $p,~\pi,~K$.  The probability for the two partons in particles $y$ and $z$ to collide is found by integrating over all possible impact parameters $\b_1$ and $\b_2$ for a given impact parameter $\b$  of the collision
\be
A_{yz}^{\rm hard} (\nu_y, \nu_z, \b) ~=~ \int d^2 {\bf{b_1}} d^2 {\bf{b_2}} \, A_y(\nu_y,{\bf{b_1}}) \, A_z(\nu_z,{\bf{b_2}}) \, \delta^{(2)}({\bf{b_1}} -
{\bf{b_2}} - \b) \ .
\ee
The profile function of a proton is given by
\ba
\displaystyle
A_p(\nu_p,\b) &\approx& \frac{1}{(2 \pi)^2} \, \int d^2 k_T \, \left(1 + \frac{k_T ^2}{\nu_p^2} \right)^{-2} \, e^{i {\bf{k_T}}\cdot \b}
\nonumber \\
& = & \nu_p \, |\nu_p \b| \, K_1(|\nu_p \b|)
\label{eq:aphard}
\ea
with $\nu_p^2 \approx 0.71$ (GeV/c$)^2$. The profile function of a meson is  
\be
\displaystyle
A_m(\nu_m,\b) ~=~ \frac{1}{(2 \pi)^2} \, \int d^2 k_T \, \left( 1 \,+\, \frac{k_T ^2}{\nu_m^2} \,+\, \eta\left(\frac{k_T^2}{\nu_m^2}\right) \,+\,\right)^{-1} \, e^{i {\bf{k}}_T \cdot \b} \ ,
\ee
where $\nu_m$ and $\eta$ are adjustable parameters. For pions, $\nu_\pi^2 \approx 0.54$ (GeV/c)$^2$ and $\eta \approx 0$. Hence the profile function for a $pp$ interaction is
\ba
\displaystyle
A_{pp}^{\rm hard}(\nu_p,\b) &=&  \int d^2 {\bf{b^\prime}} \,
A_p(|\b-{\bf{b^\prime}}|) \, A_p(|{\bf{b^\prime}}|) \nonumber \\
&=& { \nu_p^2 \over 12 \pi} {1 \over 8} \, (\nu_p b)^3 \, K_3(\nu_p b) \ ,
\label{eq:apphard}
\ea
and for a $p\pi$ interaction
\be
\displaystyle
A_{p\pi}^{\rm hard}(\nu_p,\nu_\pi,\b) ~=~ {\nu_p^2 \over 2 \pi} \, { 1
\over (1 \,-\, \zeta)} \, \left[ { \nu_p b \over 2} \, K_1(\nu_p b) \,+\, {\zeta
\over  1 \,-\, \zeta} \, [K_0(\nu_\pi b) \,-\, K_0(\nu_p b)] \right] \ ,
\ee
where $\zeta = (\nu_p / \nu_\pi)^2$. The profile functions are normalized to
\be
\int d^2 b \, A_{yz}^{\rm hard}(\nu_p,\nu_\pi,\b) ~=~ \int d^2 b \, A_y(\nu_m,\b) ~=~ 1
\ee


\section{Diffraction dissociation}
\label{app:diff}
In low-mass diffraction, the two-channel eikonal model is used \cite{Kaidalov:1979jz,Fletcher:1994hv}. Only two states are considered here; a nondiffractive state and a generic diffractive state denoted $^\star$. The diffractive scattering of particles $Y$ and $Z$ can be expressed in the following matrix elements
\begin{subequations}
\ba
\langle YZ | \mathscr{M}^{\rm int} | YZ \rangle &=&
\mathscr{M}^{\rm Born} \\
\langle YZ | \mathscr{M}^{\rm int} | Y^\star Z \rangle &=& \beta_Y \,
\mathscr{M}^{\rm Born} \\
\langle YZ | \mathscr{M}^{\rm int} | Y Z^\star \rangle &=& \beta_Z \,
\mathscr{M}^{\rm Born} \\
\langle YZ | \mathscr{M}^{\rm int} | Y^\star Z^\star \rangle &=& \beta_Y \beta_Z \,
\mathscr{M}^{\rm Born} \\
\langle Y^\star Z | \mathscr{M}^{\rm int} | Y^\star Z \rangle &=& (1 \,-\,
2\alpha_Y) \, \mathscr{M}^{\rm Born} \\
\langle Y Z^\star | \mathscr{M}^{\rm int} | Y Z^\star \rangle &=& (1 \,-\,
2\alpha_Z) \, \mathscr{M}^{\rm Born} \\
\langle Y^\star Z^\star | \mathscr{M}^{\rm int} | Y^\star Z^\star \rangle
&=& (1 \,-\, 2\alpha_Y) \, (1 \,-\, 2\alpha_Z) \,\mathscr{M}^{\rm Born} \ .
\ea
\end{subequations}
The coefficients $\alpha$ and $\beta$ may depend on energy. A matrix $\hat{\chi}(s,\b)$ for the eikonal $\chi$ is introduced to calculate $\mathscr{M}$. The eikonal matrix is diagonalized, and the cross sections calculated. The hadronic states $Y$ and $Z$ are defined as
\be
|Y,Z \rangle \,\sim\, \begin{pmatrix} 1 \\ 0 \\ 0 \\ 0 \\ \end{pmatrix} ~,~~~~
|Y^\star,Z \rangle \,\sim\, \begin{pmatrix} 0 \\ 1 \\ 0 \\ 0 \\ \end{pmatrix}
~,~~~~
|Y,Z^\star \rangle \,\sim\, \begin{pmatrix} 0 \\ 0 \\ 1 \\ 0 \\ \end{pmatrix}
~,~~~~
|Y^\star,Z^\star \rangle \,\sim\, \begin{pmatrix} 0 \\ 0 \\ 0 \\ 1 \\ 
\end{pmatrix} ~.
\ee
The eikonal matrix reads
\be
\hat{\chi}(s,\b) \,=\, 
\begin{pmatrix} 
1 & \beta_Y & \beta_Z & \beta_Y \, \beta_Z \\
\beta_Y & 1 \,-\, 2 \alpha_Y & \beta_Y \, \beta_Z & \beta_Z \, (1 \,-\, 2 \alpha_Y) \\
\beta_Z & \beta_Y \, \beta_Z & 1 \,-\, 2 \alpha_Z & \beta_Y \, (1 \,-\, 2 \alpha_Z) \\
\beta_Y \, \beta_Z ~ & \beta_Z \, (1 \,-\, 2 \alpha_Y) ~ & \beta_Y \, (1 \,-\,
2 \alpha_Z) ~& (1 \,-\, 2 \alpha_Y) \, (1 \,-\, 2 \alpha_Z)
\end{pmatrix}
\, \chi(s,\b) \ .
\ee 
After diagonalizing $\hat\chi(s,\b)$, the cross sections can be calculated. The total cross section is given by
\ba
\sigma_{YZ}^{\rm tot} &=& 2 \int d^2 \b\ \langle YZ| \left( 1 - e^{-\hat\chi(s,\b)} \right) | YZ \rangle
\nonumber \\
&=& 2 \int d^2 \b \ \sum_{n=1}^{\infty} f^{\textrm{el},n}_Y f^{\textrm{el},n}_Z (-1)^{n-1} \frac{(\chi(s,\b))^n}{n!} \ ,
\label{two-channel-tot}
\ea
where
\be
f^{\textrm{el},n}_j ~=~ (1-\frac{\alpha_j}{\gamma_j}) (1-\alpha_j-\gamma_j)^n 
\,+\, (1+\frac{\alpha_j}{\gamma_j}) (1-\alpha_j+\gamma_j)^n \ ,
\ee
and
\be
\gamma_j ~=~ \sqrt{\alpha_j^2 + \beta_j^2} \hspace*{2cm} j \,=\, Y,Z \ .
\ee
Consequently the elastic cross section reads
\be
\sigma_{YZ}^{\rm el} ~=~ \int d^2 \b \ \left| \sum_{n=1}^{\infty}
f^{\textrm{el},n}_Y f^{\textrm{el},n}_Z (-1)^{n-1}\frac{(\chi(\b,s))^n}{n!}
\right|^2 \ .
\label{two-channel-el}
\ee
The cross section for single diffraction dissociation of particle $Y$ follows from
\ba
\sigma_{YZ}^{\textrm{SD},Y} &=& \int d^2 \b\ \left| \langle Y^\star Z|  \left(
1 - e^{-\hat\chi(\b,s)} \right) | YZ \rangle \right|^2
\nonumber \\
&=& \int d^2 \b \ \left| \sum_{n=1}^{\infty} f^{\textrm{diff},n}_Y
f^{\textrm{el},n}_Z (-1)^{n-1}\frac{(\chi(\b,s) )^n}{n!} \right|^2 \ ,
\label{two-channel-sd}
\ea
using
\be
f^{\textrm{diff},n}_j ~=~ \frac{\sqrt{\gamma_j^2-\alpha_j^2}}{2\gamma_j}
\left[ (1-\alpha_j+\gamma_j)^n  \,-\, (1-\alpha_j-\gamma_j)^n  \right] \ .
\ee
Finally the expression for double diffraction dissociation is given by
\ba
\sigma_{YZ}^{\rm DD} &=& \int d^2 \b \left| \langle Y^\star Z^\star|
\left( 1 - e^{-\hat\chi(\b,s)} \right) | YZ \rangle \right|^2
\nonumber\\
&=& \int d^2 \b \left| \sum_{n=1}^{\infty} f^{\textrm{diff},n}_Y
f^{\textrm{diff},n}_Z (-1)^{n-1}\frac{(\chi(\b,s))^n}{n!} \right|^2 \ .
\label{two-channel-dd}
\ea
Note that after carrying out the sum over $n$ in Eqs.~\refb{two-channel-tot}, \refb{two-channel-el}, \refb{two-channel-sd}, \refb{two-channel-dd} the cross sections can be written as impact parameter integrals over a sum of exponentials.

The parameter range for $\alpha_j$ and $\beta_j$ is limited by the unitarity constraint that all eikonal functions have to be non-negative 
\be
1-\alpha_j-\gamma_j \ge 0 \hspace*{2cm} \alpha_j < 1/2 \hspace*{2cm} \beta_j >
0 \ .
\ee
A good description of the data is found for $\alpha = 0.2$ and $\beta = 0.5$.

The partial cross sections for $N_s$ soft and $N_h$ hard interactions follow from
\begin{equation}
\sigma_{N_s,N_h} = \int d^2b \sum_{k=1}^4 \Lambda_k \frac{[2 \lambda_k\chi_{\rm soft}(s,\b)]^{N_s}}{N_s !} \frac{[2 \lambda_k \chi_{\rm hard}(s,\b)]^{N_h}}{N_h !} 
\exp\left\{- 2\lambda_k (\chi_{\rm soft}(s, \b) + \chi_{\rm hard}(s,\b) )\right\},
\label{eq:cuts-with-diffraction}
\end{equation}
with
\begin{eqnarray}
\Lambda_1 &=& \left(1-\frac{\alpha_Y}{\gamma_Y}\right) \left(1-\frac{\alpha_Z}{\gamma_Z}\right)
\nonumber\\
\Lambda_2 &=& \left(1-\frac{\alpha_Y}{\gamma_Y}\right) \left(1+\frac{\alpha_Z}{\gamma_Z}\right)
\nonumber\\
\Lambda_3 &=& \left(1+\frac{\alpha_Y}{\gamma_Y}\right) \left(1-\frac{\alpha_Z}{\gamma_Z}\right)
\nonumber\\
\Lambda_4 &=& \left(1+\frac{\alpha_Y}{\gamma_Y}\right) \left(1+\frac{\alpha_Z}{\gamma_Z}\right)
\end{eqnarray}
and
\begin{eqnarray}
\lambda_1 &=& (1-\alpha_Y-\gamma_Y) (1-\alpha_Z-\gamma_Z)
\nonumber\\
\lambda_2 &=& (1-\alpha_Y-\gamma_Y) (1-\alpha_Z+\gamma_Z)
\nonumber\\
\lambda_3 &=& (1-\alpha_Y+\gamma_Y) (1-\alpha_Z-\gamma_Z)
\nonumber\\
\lambda_4 &=& (1-\alpha_Y+\gamma_Y) (1-\alpha_Z+\gamma_Z) \ .
\end{eqnarray}

For high-mass diffraction, it is assumed that a constant fraction of each cut (soft or hard interaction) corresponds to an rapidity-gap final state. The corresponding cross section is written, see Eq.~(\ref{eq:cuts-with-diffraction})
\begin{subequations}
\ba
\sigma_{\rm hm}^{\rm SD} &=& \delta\ (\sigma_{1,0} + \sigma_{0,1}) \ ,
\label{single-hm-diff}
\\
\sigma_{\rm hm}^{\rm DD} &=& \delta^2\ (\sigma_{1,0} + \sigma_{0,1} +
\sigma_{1,1}) \,+\, \beta_Y^2 \sigma^{\textrm{SD},Z}_{\rm lm} \,+\, 
\beta_Z^2 \sigma^{\textrm{SD},Y}_{\rm lm} \ .
\label{double-diff}
\ea
\end{subequations}
The factor $\delta$ is estimated by comparing with HERA data: 10\% of all deep inelastic scattering events at low $x$ correspond to diffraction ($\delta \approx 0.1$). The last two terms in Eq.~\refb{double-diff} represent the cross section for low-mass -- high-mass double diffraction dissociation.

\end{document}